\documentclass[%
reprint,
superscriptaddress,
longbibliography,
amsmath,amssymb,
aps,
prb,
floatfix,
]{revtex4-2}
\usepackage[T1]{fontenc}
\usepackage{graphicx}%
\usepackage{xcolor}
\usepackage{dcolumn}%
\usepackage{bm}%
\usepackage[
colorlinks,
linkcolor=blue,
citecolor=blue,
urlcolor=blue
]{hyperref}%
\usepackage[
subrefformat=parens,
labelformat=parens,
caption=false
]{subfig}
\usepackage{booktabs}
\DeclareMathOperator{\sgn}{sgn}
\begin{document}

\title{%
Anyonic analogue of optical Mach-Zehnder interferometer}
\author{Navketan Batra}\affiliation{Department of Physics, Brown University, Providence, Rhode Island 02912, USA}
\affiliation{Brown Theoretical Physics Center, Brown University, Providence, Rhode Island 02912, USA}
\affiliation{Department of Physics, University of Z{\"u}rich, 8057 Z{\"u}rich, Switzerland}
\author{Zezhu Wei}
\affiliation{Department of Materials Science and Engineering, University of Washington, Seattle, Washington 98195, USA}
\author {Smitha Vishveshwara}
\affiliation{Department of Physics, University of Illinois at Urbana-Champaign, Urbana, Illinois 61820, USA}
\author{D. E. Feldman}
\affiliation{Department of Physics, Brown University, Providence, Rhode Island 02912, USA}
\affiliation{Brown Theoretical Physics Center, Brown University, Providence, Rhode Island 02912, USA}

\date{\today}

\begin{abstract}
Anyonic interferometry is a direct probe of fractional statistics. We propose an interferometry geometry that parallels an optical Mach-Zehnder interferometer and offers several advantages over existing interferometry schemes. In contrast to the currently studied electronic Mach-Zehnder interferometer, our setup has no drain inside the device so that the trapped topological charge is time-independent. In contrast to electronic Fabry-P{\'e}rot interferometry, anyons cannot go around the device more than once. Thus, the interference signal has a straightforward interpretation in terms of anyonic statistical phases. The proposed geometry suppresses the undesirable effects of bulk-edge coupling. Moreover, the setup allows for simple exact solutions for the electric current and noise for an arbitrary quasiparticle tunneling strength in a broad range of conditions. The structure of the solutions is similar to that for non-interacting electrons but reflects fractional charge and statistics. We present results for electric current and noise in Jain states and address thermal interferometry at zero voltage bias.

\end{abstract}

\maketitle

\section{Introduction}\label{sec:intro}

Fractional statistics has been recognized \cite{Halperin1984,Arovas1984fractional} as a key feature of topological matter since the 1980s, but its direct observation proved to be a major challenge and arrived \cite{review-FH} only in the 2020s. According to the Laughlin argument, any gapped system with fractional Hall conductance has fractionally charged excitations \cite{Laughlin-arg}. A clever application of the Byers-Yang theorem \cite{BY1961}  proves that such fractional charges cannot be bosons or fermions \cite{review-FH,Kivelson1985,Hansson2025}.
Yet, the Hall conductance only sets a constraint on the type of fractional statistics, but does not fully determine it. Besides, general arguments are incapable of telling if the mutual anyonic statistics  of a small collection of fractional charges would manifest itself in any particular experiment in a particular sample. 

Several types of probes help narrow down the statistics. Thermal conductance
\cite{t1,t2,t3,texp1} is particularly useful in the search for non-Abelian anyons \cite{review-FH} since its fractional quantization is inconsistent with Abelian statistics. Thus, the observation \cite{texp2} of half-integer heat conductance in a quantum Hall liquid has brought evidence of non-Abelian quasiparticles. That evidence was indirect since thermal conductance depends on the edge physics only and is insensitive to the presence of anyons in the gapped bulk. Anyon collider experiments involve quasiparticle tunneling through the bulk \cite{rosenow2016current,bartolomei2020fractional,col2,col3}, but the interpretation of the data depends on the non-universal edge physics, which limits the information the approach can provide.

The most direct evidence of fractional statistics has come from interferometry, in which anyons are made to encircle other anyons and the observed electric current depends on the accumulated statistical phase \cite{review-FH}.
The idea \cite{chamon1997:PhysRevB.55.2331} dates to the mid-1990s, and much interesting data have been accumulated over the last two decades \cite{camino2007:PhysRevLett.98.076805,willett09:pnas.0812599106,mcclure2012:PhysRevLett.108.256804,manfra19,manfra20,nakamura2023:fabry,kundu2023:mach-zehnder,Willett2023interference,
werkmeister2025interference,Samuelson2024interference}. 
However, the first clear observation of anyonic statistics in the simplest Laughlin state at the filling factor $1/3$ arrived \cite{manfra20} only in 2020. 

\begin{figure*}
    \centering
    \includegraphics[width=0.8\linewidth]{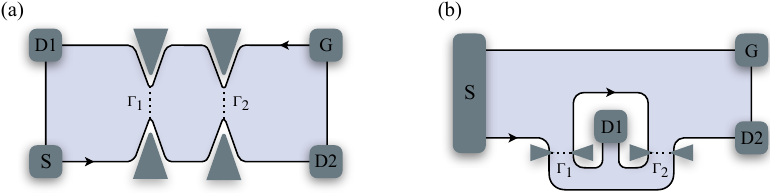}
    \caption{Previously studied anyon interferometer designs. a) A Fabry-P{\'e}rot interferometer. Charge travels along the edges from sources S and G to drains D1 and D2. Dotted lines show tunneling between the opposite edges with the amplitudes $\Gamma_{1,2}$. (b) A Mach-Zehnder interferometer. The same Ohmic contact {D1} serves as the source and drain for the inner edge. }
    \label{fig:1}
\end{figure*}

Interferometry research has focused on two geometries (Fig.~\ref{fig:1}): Fabry-P{\'e}rot \cite{chamon1997:PhysRevB.55.2331} and Mach-Zehnder \cite{MZ-heiblum}. In the limit of weak tunneling at the two point contacts, the current through a Fabry-P{\'e}rot device is a quadratic function of two tunneling amplitudes $\Gamma_{1,2}$, $I=a(\Gamma_1^2+\Gamma_2^2)+
2b\Gamma_1\Gamma_2\cos\phi$, where $a$ and $b$ are constants and the phase $\phi$ depends on the statistics of anyons. 
It also depends on the number of anyons inside the interferometer. That number can be controlled with the external magnetic field. It also depends on the shape and area of the device, which determine the number of potential minima with localized quasiparticles inside the interferometer. 
The area can be controlled with side gates. Every time a new anyon enters between the interfering paths, the phase $\phi$
jumps. The size of the jump contains information about fractional statistics.

The current behaves in a complicated way at strong and intermediate tunneling since charge is allowed to make multiple loops around the device. Hence, the Fabry-P{\'e}rot approach is most useful in the weak tunneling regime, where the experimental accuracy is limited by low signal visibility.
Mach-Zehnder geometry does not allow multiple loops, but it involves another complication: the topological charge inside the device changes after each tunneling event \cite{law2006:PhysRevB.74.045319,MZ2,feldman2007:shot_noise}. As a consequence, the relation between the statistics and observables such as electric current and noise is often complex \cite{ma2016-16}. Both geometries share another challenge due to bulk-edge interaction, which affects the interferometry phase. Indeed, any anyon inside the device interacts electrostatically with the edges and hence changes the area of the device \cite{halperin2011:PhysRevB.83.155440}. 
Thus, it affects not only the statistical phase accumulated by anyons on a path around the interferometer but also their Aharonov-Bohm phase in the external magnetic field. It is challenging to separate the two effects.

\begin{figure}
    \centering
    \includegraphics[width=0.8\linewidth]{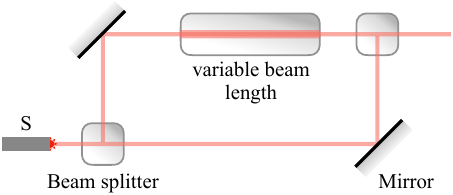}
    \caption{Optical Mach-Zehnder interferometer. The beam is emitted from source S. Once split, the beam has its paths go along the same direction and then meet at a later point. The anyonic device we present here is able to realize this key principle.}
    \label{fig:2}
\end{figure}

We propose another interferometry setup with suppressed effects of bulk-edge interactions, and which is free from other limitations of the two standard geometries. The geometry is closely related to
the optical Mach-Zehnder interferometry \cite{zehnder1891,mach1892} (Fig.~\ref{fig:2}) and differs from the existing electronic Mach-Zehnder approach by the absence of a drain inside the device. In the new geometry (Fig. \ref{fig:3}), anyons tunnel between two co-propagating channels of the same edge. Tunneling events do not change the topological charge between the channels, similar to the Fabry-P{\'e}rot interferometry. Similarly to the standard anyonic Mach-Zehnder setup \cite{MZ-heiblum}, anyons cannot make multiple loops around the device. In contrast to both setups, the charge density is different above and below the interfering channels in Fig.~\ref{fig:3}. The density assumes an intermediate value between the channels. As a result, the Coulomb interaction with any charged anyon that enters between the channels pushes the channels in the same direction. This suppresses the area change in comparison with the standard Fabry-P{\'e}rot geometry.

The setup possesses an interesting feature: an exact solution for the current and electrical noise can be found under a range of realistic conditions. The solution is remarkably simple and resembles the solution for free non-interacting electrons in the integer quantum Hall effect but incorporates the proper fractional charge and statistical phase. The origin of the exact solution can be understood from conformal field theory (CFT) \cite{difrancesco1997:conformal}. Indeed, a chiral edge is described by a chiral CFT \cite{hansson2017review}.
A tunneling operator between two edge channels enters the Hamiltonian and hence is bosonic. 
In a chiral CFT, a boson has an integer scaling dimension. In the simplest scenario, it is the same as the scaling dimension of a product of two free fermions. This opens a way to map the problem onto a system of free fermions. 

Co-propagating spin-resolved edge channels of a $\nu=2$ integer quantum Hall liquid have been theoretically investigated \cite{iyoda2025coherent} and experimentally implemented in the context of %
electronic interferometry \cite{deviatov:2008,shimizu2025mach}. In the absence of anyons, this setup is essentially equivalent to the electronic Mach-Zehnder interferometer from Ref.~\onlinecite{MZ-heiblum}. An early attempt to build an anyonic interferometer from a single quantum Hall edge involved tunneling between channels with contra-propagating modes \cite{deviatov:2009,deviatov:2012}. The setup allowed excitations to make multiple loops around the device and was thus similar to a Fabry-P{\'e}rot interferometer. No signatures of anyonic statistics were observed in Refs.~\onlinecite{deviatov:2009,deviatov:2012}.
Such signatures were revealed in very recent experiments
\cite{chiral2024-1,chiral2024-2}, which followed the theoretical proposal \cite{wei:2023-chiral}.

This article follows the Letter \cite{wei:2023-chiral}, which focused on statistics in the Laughlin $\nu=1/3$ state. 
The setup involved tunneling between two edge channels of the $\nu=2/5$ quantum Hall liquid. 
This filling factor is relevant for the recently discovered quantum anomalous Hall effect in 
MoTe$_2$, where $\nu=-2/3$ and $\nu=-3/5$ have been observed \cite{cai2023,zheng2023,park2023,PhysRevX.13.031037}. In this paper, we extend the results of Ref.~\onlinecite{wei:2023-chiral} to an arbitrary Jain state. The exact solutions for the electric current and noise are possible in two limits: 1) low voltage and temperature and 2) equal propagation times of charge between the tunneling contacts in both channels. To understand what changes when these conditions are violated, we use perturbation theory to compute the current beyond the exactly solvable cases. 
Heat transport proved to be a powerful probe of anyonic statistics.
In this paper, we address thermal interferometry \cite{therm-int-1,therm-int-2,therm-int-3} where the two adjacent co-propagating modes are maintained at different temperatures to facilitate thermal current. 
Like the electric current, the thermal current depends on the mutual statistical phase of anyons.

The paper is organized as follows. We begin in Section \ref{sec:models} by introducing the interferometer geometry and the setup for electrical and thermal transport. 
Thermal interferometry \cite{therm-int-1,therm-int-2,therm-int-3} gives information about anyonic statistics and it is instructive to compare it with interferometry of charge currents. Then, in Section \ref{sec:electric_transport}, we use perturbation theory to compute the average electric current in the most general case.
Here we observe the key feature that the tunneling quasiparticle operator has the same scaling dimension as the free-fermion tunneling operator in the integer quantum Hall effect (IQHE), and in Section \ref{sec:exactsol}, we find an exact solution for the electric current by mapping the problem to free-fermion scattering. Next, in Section \ref{sec:thermal_transport}, we extend the calculations to the thermal current using the machinery developed before. This is rather technical and heavily relies on the Appendix. The exact solution does not work in the thermal transport problem.
Finally, in Section \ref{sec:noise}, we compute the electrical noise before concluding in Section \ref{sec:conclusions}. Details regarding the calculations can be found in the Appendices. A detailed list of all the notations and variables used in this paper can be found in Appendix \ref{appendix:notations}. We summarize our main results below.

\subsection{Main results}

We find exact solutions for the electric current and noise. The solutions apply in two limits: 1) low voltages and temperatures or 2) equal travel times between the two constrictions along the two channels. Regardless of the tunneling strength between the two channels, the current has a simple form of 
\begin{equation}
\label{EqI}
I=A+B\cos\varphi, 
\end{equation}
where $A$ and $B$ are constants, and the phase $\varphi$ jumps by the mutual statistical phase of two anyons when each new anyon enters the device. For the Jain state at the filling factor $k/(2k+1)$, the phase jump is $-4\pi/(2k+1)$. The phase also contains an Aharonov-Bohm contribution, which is proportional to the quasiparticle charge.  The noise is found to have a simple form with two nonzero harmonics of the statistical phase.

Beyond the limits 1) and 2), we compute the electrical current perturbatively in the tunneling amplitudes. Within perturbation theory, the current always follows form (\ref{EqI}). In the same approximation, noise can be expressed in terms of the current with the help of the detailed balance principle. We also compute in the lowest-order perturbation theory the thermal current between the two channels at zero voltage bias under the assumption that the two interfering channels emanate from Ohmic contacts with different temperatures. 
The heat current again has the general form (\ref{EqI}).

\section{Models}\label{sec:models}

\begin{figure}
    \centering
    \includegraphics[width=0.9\linewidth]{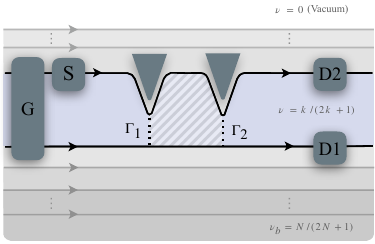}
    \caption{An anyonic analogue of an optical Mach-Zehnder interferometer. Arrows on thick black lines show the direction of charge propagation. Two constrictions allow quasiparticle tunneling between the two channels with the amplitudes $\Gamma_{1,2}$.}
    \label{fig:3}
\end{figure}

We consider an edge of a topological liquid whose bulk is in a fractional quantum Hall (FQH) state with the filling factor $\nu_b=N/(2N+1)$. The edge structure consists of $N$ co-propagating modes \cite{WenBook}. We introduce the label $k=1,\dots, N-1$ which defines the edge mode with the conductance $e^2\nu_1/h=e^2/[(2k-1)(2k+1)h]$ that separates the incompressible regions of filling factors $\nu=(k-1)/(2k-1)$ and $\nu=k/(2k+1)$. Similarly, the inner adjacent edge mode of conductance $e^2\nu_2/h=e^2/[(2k+1)(2k+3)h]$ separates the incompressible regions of $\nu=k/(2k+1)$ and $\nu=(k+1)/(2k+3)$ as shown in Fig.~\ref{fig:3}. The two adjacent co-propagating modes are brought nearby at some points with external gates  to facilitate quantum tunneling through constrictions. Since the filling factor of the incompressible region separated by the two modes is $\nu=k/(2k+1)$, the fundamental quasiparticle that tunnels between the two co-propagating modes has a charge of $e^*=e/(2k+1)$. Multiple constrictions define the interferometer geometry, and for simplicity, we consider an interferometer with two constrictions in this work.  

Several remarks on the geometry are in order. Since the edge modes co-propagate between the two constrictions, the above geometry is a variation of a Mach-Zehnder interferometer. Additionally, we see that, in contrast to the Fabry-P{\'e}rot geometry, due to their travel along co-propagating channels, anyons cannot make multiple loops in the device on the path from source to drain. Furthermore, fabrication of the geometry does not require an Ohmic contact inside the device, so that the topological charge in the device remains unchanged after each tunneling event, contrary to a conventional Mach-Zehnder device. These features make the device particularly interesting and motivate this work. 

In the bosonization language \cite{vondelft1998:bosonization}, we introduce the Bose fields $\phi_1$ and $\phi_2$ which model the adjacent co-propagating edge modes. The Lagrangian density of these two right-moving edge modes \cite{WenBook} is 
\begin{align}
\mathcal{L}&=-\frac{\hbar}{4\pi} \sum_{i=1,2} \left[ \frac{1}{\nu_i}\partial_t\phi_i \partial_x \phi_i + \frac{v_i}{\nu_i}(\partial_x\phi_i)^2 \right].
\end{align}
The subscripts $1$ and $2$ refer to the outer and inner adjacent channels, respectively, with the mode velocities $v_1$ and $v_2$ and the effective filling factors $\nu_1=1/[(2k-1)(2k+1)]$ and $\nu_2=1/[(2k+1)(2k+3)]$. The Bose fields $\phi_i$ obey the commutation relation 
\begin{align}
[\phi_i(x),\phi_j(x')]=i\pi \delta_{ij}\nu_i\sgn{(x-x')}.
\end{align}
The charge density on the $i$th edge mode is then defined as $\rho_i=e\partial_x \phi_i/2\pi$. 
The correlation functions of the fields $\phi_i$ at a non-zero temperature $\Theta$ is given as
\begin{align}
\label{cor-fun}
&\langle \phi_i(x,t)\phi_j(0,0) \rangle = \\
&-\delta_{ij} \nu_i\text{ln}\left[ \frac{2\pi}{L} \frac{v_i\hbar}{\pi k_B \Theta} \sin\left( \frac{\pi k_B \Theta}{\hbar}[\epsilon+ i (t-x/v_i)] \right)  \right], \nonumber
\end{align}
where $\epsilon$ is a short-time cutoff and $L$ is the length of the edge, which we send to infinity.
The above action ignores the Coulomb interaction between the edge channels since it is screened in the presence of the top gates. Screening is less efficient near the tunneling contacts between the two edge channels, but Coulomb interaction in those regions is irrelevant in the renormalization group sense at low energies.

Since the filling factor of the incompressible region separated by the two modes is $\nu=k/(2k+1)$, the fundamental quasiparticle that tunnels between the two co-propagating modes has a charge of $e^*=e/(2k+1)$. An $e^*$ quasiparticle on the outer channel is created by
\begin{align}
    \Psi_1^{\dagger} = \left( \frac{L}{2\pi} \right)^{-\alpha/2} F_1^{\dagger} : e^{-i(2k-1)\phi_1}:, 
\end{align}
where $\alpha = (2k-1)/(2k+1)$ and the rest of the notation is explained below. Similarly, a quasiparticle of charge $e^*$ on the inner edge channel is created by
\begin{align}
    \Psi_2^{\dagger} = \left( \frac{L}{2\pi} \right)^{-\beta/2} F_2^{\dagger} : e^{-i(2k+3)\phi_2}:,
\end{align}
where $\beta = (2k+3)/(2k+1)$. Here, :~: represents normal ordering and the operators $F_{1,2}$ are the Klein factors \cite{vondelft1998:bosonization}. The exponents $\alpha$ and $\beta$ reflect the scaling dimension of the quasiparticle operators, and interestingly, their sum $\alpha+\beta=[(2k-1)/(2k+1)] + [(2k+3)/(2k+1)]=2$ for any two adjacent incompressible modes defined by the label $k=1,\dots,N-1$ for an incompressible bulk at filling $\nu_b=N/(2N+1)$. Following this, we define the tunneling operators that transfer a quasiparticle of charge $e^*=e/(2k+1)$ from the outer edge at a point $x$ to the inner edge at a point $y$ as,
\begin{align}
T(x,y) &= \Psi_2^{\dagger}(y)\Psi_1(x) \\
& =  \left( \frac{L}{2\pi} \right)^{-1} F_2^{\dagger} F^{}_1 : e^{-i(2k+3)\phi_2(y)}: : e^{i(2k-1)\phi_1(x)}:. \nonumber
\end{align}
We use different coordinates on the two edge channels because, in general, they have different lengths between the two tunneling contacts. We will set the coordinates of the left contact to $0$ in both channels. The coordinates of the right constriction are $x=l+a$ and $y=l$.
With these definitions, the commutator of the tunneling operator with the total charge, defined as $Q_i \equiv \int dx \rho_i$ on the $i$th channel, results in $\left[ Q_1,T(x,y) \right] = -e^* T(x,y)$ and $\left[ Q_2,T(x,y) \right] = e^* T(x,y)$, restating the transfer of an $e^*$ quasiparticle. 

\subsection{Setup for electric transport}\label{sec:models_setup_electric}

The Hamiltonian of the interferometer comprises two components, the edge-mode Hamiltonian $H_0$ and the tunneling Hamiltonian $H_T$,
which contains two tunneling operators $T(0,0)$ and $T(l+a,l)$.
Here, $l$ and $l+a$ are the distances between the tunneling contacts along the two edge channels. 
The total Hamiltonian $H = H_0 + H_T$ can be written as
\begin{align}
H &= \pi \hbar \int_{-L/2}^{L/2} dx \left[ \frac{v_1}{\nu_1 e^2} \rho_1^2 + \frac{v_2}{\nu_2 e^2} \rho_2^2  \right] \nonumber\\  
&~~~+ \left[ \Gamma_1 T(0,0) + e^{i\varphi} \Gamma_2 T(l+a,l) + \text{H.c.} \right].
\end{align}
where the free edge-mode Hamiltonian $H_0$ and the tunneling Hamiltonian $H_T$ are defined as
\begin{align}
\label{9-new}
H_0 &= \pi \hbar \int_{-L/2}^{L/2} dx \left[ \frac{v_1}{\nu_1 e^2} \rho_1^2 + \frac{v_2}{\nu_2 e^2} \rho_2^2  \right], \\  
\label{10-new}
H_T &= \Gamma_1 T(0,0) + e^{i\varphi} \Gamma_2 T(l+a,l) + \text{H.c.} 
\end{align}%
Here we introduced a phase $\varphi$, which combines the Aharonov-Bohm phase, the statistical phase accumulated by a tunneling anyon on the path around quasiparticles confined between the edge channels, and a non-universal phase difference between the two constrictions. 

A change in the magnetic field or a gate voltage
changes the number of confined anyons. When that number changes,
the phase $\varphi$ jumps. The size of the jump \cite{KF1997} is $\Delta\varphi =  -4\pi/(2k+1)$, for $k=1,\dots,N-1$. Fractional statistics can be determined from the knowledge of these jumps. Coulomb interaction of confined anyons with the edges of the devices shifts the edges and hence affects the Aharonov-Bohm phase. Even a tiny shift of the edge might be sufficient to change $\varphi$ on the order of 1. 
This negatively impacts the accuracy of the probe of the fractional statistics. The issue is common to all interferometry setups, but it is less important in the setup of Fig.~\ref{fig:3} than in the standard Fabry-P{\'e}rot geometry. Indeed, consider, for example, an anyon that attracts a negative charge and hence increases the filling factor near the edge channels. Since the filling factor grows from the bottom to the top of the device in Fig.~\ref{fig:3}, both edge channels shift down. These shifts have opposite effects on the Aharonov-Bohm phase. In contrast, the edges of a Fabry-P{\'e}rot interferometer shift in the opposite directions.

Besides a single quasiparticle,
multiple quasiparticle charges 
are allowed to tunnel between the edges. We ignore such processes since the corresponding tunneling operators are less relevant than $T$.

In the electric transport measurement,  the two edge channels are connected to the sources at the potentials $V_1$ and $V_2$. We will assume that $e(V_1-V_2)\equiv eV>0$. In the interaction representation \cite{law2006:PhysRevB.74.045319},
we subtract the products of the chemical potentials and the charges of the edge channels from the Hamiltonian.
The edge-mode Hamiltonian in the interaction representation is
\begin{align}
K_0 = H_0 - \int_{-L/2}^{L/2} dx \left[ V_1\rho_1 + V_2 \rho_2 \right],
\end{align}
where $\rho_i = e\partial_x\phi_i/2\pi$ is the electron density in channel $i=1,2$. 
Using Hamilton's equations of motion, we find the tunneling current operator $I_T$, which in the interaction representation is given as
\begin{align}
    I_T = \frac{ie^*}{\hbar}\Big[  \Gamma_1 T(0,0;t) + e^{i\varphi} \Gamma_2T(l+a,l;t) - \text{H.c.}\Big] .
\end{align} 
In the interaction representation, the tunneling operator $T$ gets multiplied with a phase $e^{-i\omega_q t}$, where $\omega_q = e^* V/\hbar$:
\begin{align}
T(x,y) \rightarrow e^{-i\omega_q t} T(x,y;t) .
\end{align}
Additionally, we note that the difference in the chemical potentials results in different average densities in the modes. The mean charge densities in the edge modes can be found by minimizing the Hamiltonian $K_0$ with respect to the charge densities $\rho_i$. This yields $\langle \rho_i\rangle = \nu_i e^{2}V_i/2\pi\hbar v_i$. By shifting the fields $\phi_i \rightarrow \phi_i + (e\nu_i V_i /\hbar v_i)x$, we set the average density to zero for both modes at the price of introducing a phase factor in the tunneling operator,
\begin{align}
\label{time-factor}
T(x,y) \rightarrow T(x,y) \exp\left[ i \frac{e^*}{\hbar} \left( \frac{V_1 x}{v_1} - \frac{V_2 y}{v_2} \right)  \right].
\end{align} 
We denote the voltage-induced phase $\theta(x,y) \equiv e^*(V_1x/v_1 - V_2 y/v_2)/\hbar$. After taking into account the effect of different chemical potentials on the two channels, the tunneling operators in the interaction representation are given by
\begin{align}
    T(x,y) \rightarrow e^{-i\omega_q t} e^{i\theta(x,y)} T(x,y;t).
\end{align}
With these transformations, we will perturbatively (in tunneling amplitudes $|\Gamma_i|$) compute the electric current, and subsequently, obtain an exact solution for the electric current and noise.
In the next section, we introduce a setup for thermal transport. Maintaining the two edge modes at the same chemical potential results in no tunneling electric current; however, a non-equilibrium thermal current can flow if the two modes have different temperatures. We do not consider thermoelectric effects in this work.

\subsection{Setup for thermal transport}\label{sec:models_setup_thermal}

The tunneling quasiparticles between the $\phi_1$ and $\phi_2$ modes carry both charge and energy; therefore, if we keep the two sources at different temperatures, say $\Theta_1$ and $\Theta_2$, then we obtain a thermal current through the interferometer. 
Hence, the correlation functions (\ref{cor-fun}) for the fields $\phi_1$ and $\phi_2$ are defined at the temperatures $\Theta_1$ and $\Theta_2$ respectively. Notice that the phases we introduced to the tunneling operators due to the shift in the average densities in the two modes and the interaction picture treatment do not appear in this case, as the two modes are maintained at the same chemical potential. For simplicity, we consider $V_1=V_2=0$. 

To find the heat current operator $J_Q$, we notice that the energy transport can be defined using the Heisenberg equation of motion,
\begin{align}
    J_Q \equiv -\frac{d}{dt} H_2 = \frac{i}{\hbar}\left[ H_2,H \right], 
\end{align}
and $H_2=\frac{\pi\hbar v_2}{e^2\nu_2}\int dx \rho_2^2$ is the Hamiltonian of the $\phi_2$ mode.
Thus,
\begin{align}\label{eq:heat_current_op}
J_{Q}(t) &\equiv \left[ \Gamma_1 \left( \partial_t \Psi_2^{\dagger}(0,t) \right) \Psi_1(0,t) +\text{H.c.} \right]  \\
&~~~+ \left[ \Gamma_2 e^{i\varphi} \left( \partial_t \Psi_2^{\dagger}(l,t) \right) \Psi_1(l+a,t) + \text{H.c.} \right]. \nonumber
\end{align} 
We again introduced a phase $\varphi$, which consists of the Aharonov-Bohm phase, the statistical phase, and a non-universal phase difference between the two constrictions. As seen from the definition of the thermal current operator, we will be working with the derivatives of the tunneling operators. Thus, for convenience, we define
\begin{align}
T'(x,y;t) \equiv \left( \partial_t\Psi_2^{\dagger}(y,t) \right) \Psi_1(x,t) .
\end{align}
Notice the unusual notation, for the temporal derivative only acts on the quasiparticle operator $\Psi_2$.

As in the electric transport case, we ignore the contribution from the tunneling of multiples of a quasiparticle charge. Heat can be transferred by the operator $\partial_x\phi_1(0)\partial_x\phi_2(0)$ that transfers no charge. We ignore it too. As we will see from the calculation of the scaling dimension of the tunneling operator $T$, this contribution is less relevant than the operator $T$ at low energies.

\section{Electric transport}\label{sec:electric_transport}

Maintaining the two adjacent co-propagating edge modes at different chemical potentials results in a current flowing at a quantum point contact. In the conventional setups where the quantum point contact brings the two counter-propagating edges close to each other, the fundamental quasiparticles associated with the bulk filling factor dominate the current-voltage response only in the weak tunneling limit. Indeed, in the strong tunneling limit, one has to switch to the dual geometry, where the dominant tunneling particles are electrons. In contrast, the present setup with constrictions facilitating tunneling between two adjacent co-propagating modes on the same edge of the FQH liquid should be understood in terms of quasiparticle tunneling for all tunneling strengths. This occurs because particles cannot make multiple loops in the device. In this section, 
we start by computing the tunneling current in the second-order perturbation theory. We will assume that the charge propagation time between the two constrictions is shorter along the outer edge. Calculations are essentially the same in the opposite case.
We will also consider the limit of equal propagation times.

\subsection{Single constriction geometry}\label{sec:electric_transport_single}

As a warm-up exercise, we first consider the effect of a single constriction on the average tunneling current. In this case, the quasiparticle statistics do not play any role. 
Let us assume $\Gamma_2 = 0$ while $\Gamma_1= \Gamma$.
We can assume that $\Gamma_1$ is real and positive, $\Gamma_1=|\Gamma_1|$, without
loss of generality.
Then one finds the tunneling current operator in the interaction picture as
\begin{align}
I_T(t) = \frac{ie^*\Gamma}{\hbar}\Big[  e^{-i\omega_q t}  T(0,0;t) - e^{i\omega_q t}  T^{\dagger}(0,0;t) \Big] .
\end{align}  
Since the average of the tunneling current at equilibrium is zero, $\langle T \rangle_0 = 0$, the first non-zero contribution to the tunneling current comes from
\begin{align}
    \langle I_T(t) \rangle^{\text{non-int}}_{\Gamma} &= -\frac{i}{\hbar} \int_{-\infty}^{t} dt' \langle \left[ I_T(t),H_T(t') \right] \rangle_0 ,
\end{align}
where the operators on the right-hand side are expressed in the interaction representation. Plugging in the definition of the tunneling operators with only a single constriction, the result can be expressed in terms of the Fourier transforms
\begin{align}
   \langle I_T(t) \rangle^{\text{non-int}}_{\Gamma} &=  \frac{e^*|\Gamma|^2}{\hbar^2} \left[ \tilde G^>(\omega_q) - \tilde G^<(\omega_q) \right],
\end{align}
where we defined the Fourier transform of the tunneling correlation functions as $\tilde G^>(\omega_q) = \tilde G^{<}(-\omega_q) \equiv \int d\eta e^{-i\omega_q \eta}\langle T(\eta)T^{\dagger}(0) \rangle_0 $. Note that the average $\langle \cdot \rangle_0$ is computed with respect to the unperturbed thermal state. The Fourier transforms are computed in Appendix \ref{appendix:I_special}. The resulting average tunneling (non-interference) current is
\begin{align}
    \langle I_T \rangle^{\text{non-int}}_{\Gamma} = - \frac{1}{(2k+1)^2} \frac{e^2}{h} V \left( \frac{2\pi |\Gamma|}{v_1^{\alpha/2}v_2^{\beta/2}\hbar} \right)^2,
\end{align}
where the exponents $\alpha = (2k-1)/(2k+1)$ and $\beta = (2k+3)/(2k+1)$. Interestingly, the first non-zero contribution to the tunneling current follows a linear $I$-$V$ relation independently of temperature $\Theta$, similar to the case of electron tunneling in IQHE. Next, we move on to add the second constriction.
This will allow the incorporation of the effects of quasiparticle statistics.

\subsection{Double constriction geometry}\label{sec:electric_transport_double}
We now work with two constrictions and treat $\Gamma_{1}$ and $\Gamma_2$ perturbatively. The electric tunneling current now comprises the non-interference and interference contributions. As introduced earlier, the current operator in the interaction picture is given as
\begin{align}
I_T(t) &= \frac{ie^*}{\hbar}\Big[ e^{-i\omega_q t} |\Gamma_1| T(0,0;t) - \text{H.c.} \Big]\\ 
&~~~+\frac{ie^*}{\hbar} \Big[ e^{-i\omega_q t} e^{i[\varphi+\theta(l+a,l)]} |\Gamma_2|T(l+a,l;t) - \text{H.c.}\Big], \nonumber
\end{align}
where  $\text{H.c.}$ represents the Hermitian conjugate. For convenience, we introduce the notation, $T(0,0;t)\equiv T_1(t)$ and $T(l+a,l;t)\equiv T_2(t)$. We absorb all the phase factors into $\Gamma_{1,2}$, and thus we need to remember that the tunneling amplitudes are complex. Another point to remember is that since the phase factors are time-dependent, the tunneling amplitudes consequently also carry time dependence. 

The first non-zero correction to the tunneling current is computed as $\langle I_T(t)  \rangle = -(i/\hbar) \int_{-\infty}^{t} dt' \langle \left[ I_T(t),H_T(t') \right] \rangle_0$. After simplifying, one can express the total tunneling current as
\begin{align}
    \langle I_T(t)  \rangle &= \langle I_T(t) \rangle^{\text{int}} + \sum_{i=1}^2 \langle I_T(t) \rangle^{\text{non-int}}_{\Gamma_i},
\end{align}
where the non-interference contribution from each constriction, $ \langle I_T \rangle^{\text{non-int}}_{\Gamma_i} $, is simply the single constriction result, computed in Section \ref{sec:electric_transport_single}. Thus,
\begin{align}
 \langle I_T \rangle^{\text{non-int}}_{\Gamma_i} &= - \frac{1}{(2k+1)^2} \frac{e^2}{h} V \left( \frac{2\pi |\Gamma_i|}{v_1^{\alpha/2}v_2^{\beta/2}\hbar} \right)^2.
\end{align}
We now focus on the interference contribution $\langle I_T\rangle^{\text{int}}$. It can be expressed as
\begin{align}
    \langle I_T(t) \rangle^{\text{int}}  &= \frac{e^*|\Gamma_1\Gamma_2|}{\hbar^2}  \Bigg[ \int^{\infty}_{-\infty} d\eta e^{-i\omega_q \eta} e^{-i\phi}  \langle \big[ T_1(\eta) , T_2^{\dagger}(0) \big] \rangle_0 \nonumber \\
    &~~~+ \int_{-\infty}^{\infty} d\eta e^{-i\omega_q \eta} e^{i\phi} \langle  \big[ T_2(\eta),T_1^{\dagger}(0)  \big] \rangle_0  \Bigg],
\end{align}
where, for convenience, we define the phase $\phi\equiv \varphi + \theta(l+a,l)$. Similarly to the non-interference current, one can express the interference contribution in terms of the Fourier transform of the correlators $G^>(x,y;t)=\langle T(x,y;t)T^{\dagger}(0,0;0) \rangle_0$ and $G^<(x,y;t)=\langle T^{\dagger}(0,0;0)T(x,y;t) \rangle_0$, which are defined as
\begin{align}
    G^>(x,y;t) = \left[ \mathcal{G}^>_1(x,t) \right]^{\alpha}  \left[  \mathcal{G}^>_2(y,t) \right]^{\beta},\\
    G^<(x,y;t) = \left[ \mathcal{G}^<_1(x,t) \right]^{\alpha}  \left[  \mathcal{G}^<_2(y,t) \right]^{\beta}.
\end{align}
These are  Green's functions at a non-zero temperature $\Theta$ defined in terms of the functions $\mathcal{G}^{>}_i(x,t)$ and $\mathcal{G}^{<}_i(x,t)$,
\begin{align}\label{eq:thermal_greens_functions}
    \mathcal{G}^{>}_i(x,t) = \frac{\pi k_B \Theta_i}{v_i \hbar \sin\left[ \frac{\pi k_B \Theta_i}{\hbar} \left( \epsilon + i (t-x/v_i) \right) \right]},\\
    \label{eq:t-g-f-2}
    \mathcal{G}^{<}_i(x,t) = \frac{\pi k_B \Theta_i}{v_i \hbar \sin\left[ \frac{\pi k_B \Theta_i}{\hbar} \left( \epsilon - i (t-x/v_i) \right) \right]},
\end{align}
where the index $i=1,2$ represents the two edge modes with the velocities $v_i$. Notice that the mode index $i$ also appears in the subscript of temperature for the case of the two modes maintained at different temperatures. In this section, we set $\Theta_1=\Theta_2=\Theta$. When we deal with thermal transport, we will allow these temperatures to be different. These definitions are used to express the interference contribution as
\begin{widetext}
\begin{align}
\langle I_T \rangle^{\text{int}}  = \frac{e^*|\Gamma_1\Gamma_2|}{\hbar^2}  \Bigg[& e^{-i\phi}  \big[ \tilde G^>(-l-a,-l;\omega_q) - \tilde G^<(-l-a,-l;\omega_q) \big]  + e^{i\phi}  \big[ \tilde G^>(l+a,l;\omega_q) - \tilde G^<(l+a,l;\omega_q) \big]   \Bigg],
\end{align}
where $\tilde G^>(x,y;\omega) = \int dt e^{-i\omega t} G^>(x,y;t)$ is the Fourier transform of Green's function, and similarly for $G^<(x,y;t)$. To proceed, we change the integration variable to a dimensionless quantity $\tau \equiv \pi k_B \Theta t/\hbar$. We also define $\tau_1 \equiv \pi k_B \Theta (l+a)/\hbar v_1$ and $\tau_2 \equiv \pi k_B \Theta l/\hbar v_2$ as the dimensionless propagation times for the two modes to travel from one constriction to the other. These definitions simplify the above expression to
\begin{align}
\label{Eq30}
   \langle I_T \rangle^{\text{int}}  &= -2i\frac{e^*|\Gamma_1\Gamma_2|}{\hbar^2} \frac{(\pi k_B \Theta/\hbar)}{ v_{1}^{\alpha} v_{2}^{\beta} }  \Bigg[ \int_{-\infty}^{\infty}d\tau  \sin (\lambda \tau-\phi') \sin^{-\alpha}(\epsilon+i\tau) \sin^{-\beta}\left[\epsilon+i(\tau-\chi)\right] \nonumber \\
& ~~~~~~~~~~~~~~~~~~~~~~~~~~~~~~~ - \int_{-\infty}^{\infty}d\tau  \sin(\lambda \tau - \phi') \sin^{-\alpha}(\epsilon-i\tau) \sin^{-\beta}\left[\epsilon-i(\tau-\chi)\right] \Bigg] ,
\end{align}
\end{widetext}
where we defined $\lambda \equiv \hbar \omega_q /\pi k_B \Theta$, $\phi'\equiv \phi-\lambda \tau_1$, and $\chi\equiv \tau_2-\tau_1$. The variable $\chi$ represents the dimensionless propagation  time difference between the two co-propagating modes.
We assume below that $\chi\ge 0$. Calculations are similar in the opposite case.
In the limit $\chi\rightarrow 0$, the above integral becomes straightforward, as we show next. 

\subsubsection{Equal propagation times}\label{sec:electric_transport_double_equal}
In the case of equal propagation times, {\it i.e.}, when $\tau_1 = \tau_2$, or $\chi\rightarrow 0$, one immediate observation is that $\theta(l+a,l)=\lambda \tau_1$, and hence $\lambda\tau - \phi' = -\varphi + \lambda\tau$. From this observation and $\alpha + \beta = 2$, we find that the integral
\begin{align}
    &\lim_{\chi\rightarrow 0} \int_{-\infty}^{\infty}d\tau \frac{\sin \left(\lambda \tau-\phi'\right)}{\sin^{\alpha}(\epsilon \pm i\tau) \sin^{\beta}\left[\epsilon \pm i(\tau-\chi)\right]} \nonumber \\
    ={}& \frac{1}{2i} \int_{-\infty}^{\infty}d\tau \big( e^{-i\varphi}e^{\pm i\lambda \tau}  - e^{i\varphi}e^{\mp i\lambda \tau}  \big) \sin^{-2}(\epsilon+i\tau) .
\end{align}
Taking the difference of the two integrals results in the following expression for the interference contribution:
\begin{align}
    \lim_{\chi\rightarrow 0} \langle I_T \rangle^{\text{int}}=-2\cos\varphi & \frac{e^*|\Gamma_1\Gamma_2|}{\hbar^2} \frac{(\pi k_B \Theta/\hbar)}{ v_{1}^{\alpha} v_{2}^{\beta} } \nonumber \\
    & \times \int_{-\infty}^{\infty}d\tau \frac{ e^{i\lambda \tau} - e^{-i\lambda \tau} }{\sin^{2}\left(\epsilon+i\tau\right)}.
\end{align}
Therefore, we see that in the limit of equal propagation times, and assuming identical constrictions with $|\Gamma_i|=\Gamma$ for $i=1,2$, we have $\lim_{\chi\rightarrow 0} \langle I_T \rangle^{\text{int}} = 2\cos\varphi ~ \langle I_T \rangle^{\text{non-int}}_{\Gamma}$. Thus, the interference contribution can be expressed in terms of the non-interference contribution modulated with the quasiparticle statistics phase. Therefore, when $\Gamma_i=\Gamma$ for $i=1,2$, in the case of equal propagation times, the total tunneling current simply becomes
\begin{align}
   \lim_{\chi\rightarrow 0} \langle  I_T \rangle = 2(1+\cos\varphi) \langle I_T \rangle^{\text{non-int}}_{\Gamma},
\end{align}
where $\langle I_T\rangle$ is linear in voltage. This simple linear $I$-$V$ relation is a feature of the special Mach-Zehnder geometry of the device.
The total tunneling current is sensitive to the quasiparticle charge $e^*=e/(2k+1)$ and quasiparticle statistics through its $1+\cos\varphi$ dependence. Recall that $\varphi$ jumps
by $-4\pi/(2k+1)$ with each new quasiparticle in the device.  Next we focus on the general case when $\chi\ne 0$. 

\subsubsection{Arbitrary propagation times }
\label{sec:electric_transport_double_general}

\begin{figure*}
    \centering
    \includegraphics[width=0.8\linewidth]{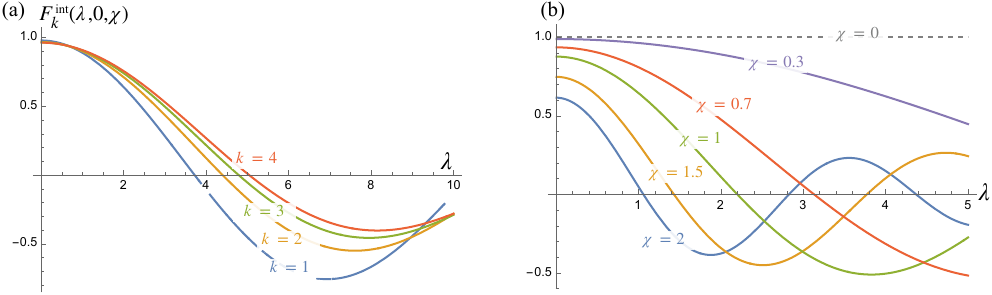}
    \caption{
    We consider $\varphi=0$, $V_2=0$, and the filling factor of the topological liquid probed by the device is $k/(2k+1)$. Panel (a) shows the dependence of $F^{\rm int}_k$ on the normalized bias $\lambda$ at a  fixed difference $\chi=0.5$ of the normalized propagation times between the constrictions along the two channels.   Panel (b) shows the dependence of $F^{\rm int}_2$, which corresponds to a fixed filling factor $\nu=2/5$, on the bias $\lambda$ at various differences $\chi$ of the travel times along the two paths.  In the limit $\chi\rightarrow 0$,  a simple result (\ref{Eq36}) is recovered. Beyond that limit, the interference contribution to the current oscillates as the voltage bias increases.}
    \label{fig:4}
\end{figure*}

We now focus on the general case where the propagation times in both modes, in general, may be different. For this, we need to solve the integrals in the expression (\ref{Eq30}) above, for the case when $\alpha+\beta=2$ with $\alpha,\beta\in \mathbb{Q}^+$. Notice that since $\alpha= (2k-1)/(2k+1)$ and $\beta=(2k+3)/(2k+1)$, we also get $\alpha<\beta$, which will be useful. In Appendix \ref{appendix:I_general}, we compute the integrals in the general case. This results in the following expression for the interference contribution to the tunneling current:
\begin{align}
\label{inter-1}
    \langle I_T \rangle^{\text{int}} &= -\frac{2}{(2k+1)^2}\frac{e^2}{h} V \left( \frac{2\pi }{v_1^{\alpha/2}v_2^{\beta/2}\hbar} \right)^2 |\Gamma_1 \Gamma_2|  \\
    &\times F_{k}^{\text{int}}\left[ \frac{e^* V_1}{\pi k_B \Theta }, \frac{e^* V_2}{\pi k_B \Theta},\chi \right], \nonumber
\end{align}
where the variable $\chi\equiv \left[l/v_2 - (l+a)/v_1\right]\pi k_B \Theta/\hbar$ is a dimensionless measure of the propagation time difference between adjacent modes across the two constrictions. In the above expression for the average interference current, the function $F_{k}^{\text{int}}(\lambda_1,\lambda_2,\chi)$ is defined in terms of the integral
\begin{widetext}
\begin{align}
\label{inter-2}
    F_{k}^{\text{int}}(\lambda_1,\lambda_2,\chi) &= \frac{\sin\left( \frac{2\pi}{2k+1} \right)}{\pi (\lambda_1-\lambda_2)} \Bigg[ \int_{0}^{\chi} d\tau \left( \frac{\sin[\varphi - \lambda_2 \chi - (\lambda_1-\lambda_2)\tau]}{\sinh^{\frac{2k-1}{2k+1}}\tau \sinh^{\frac{2k+3}{2k+1}}(\chi-\tau)} - \frac{\sin(\varphi - \lambda_1 \chi)}{(\sinh^{\frac{2k-1}{2k+1}}\chi)(\chi-\tau)^{\frac{2k+3}{2k+1}}} \right) \nonumber \\
    &~~~~~~~~~~~~~~~~~~~~~~~~~~~~~~~~~~~~~~~~~~~~~~~~~~~~~-  \frac{2k+1}{2\sinh^{\frac{2k-1}{2k+1}}\chi}\frac{\sin(\varphi-\lambda_1\chi)}{\chi^{\frac{2}{2k+1}}} \Bigg],
\end{align}
\end{widetext}
and where we also used the relation $\phi'-\lambda\tau = \varphi - \lambda_2 \chi - (\lambda_1-\lambda_2)\tau $. Note that the above function depends on the electric potentials $V_{1,2}$  of the two modes. However, in the $\chi\rightarrow 0$ limit, we see that only the potential difference matters. It can be checked (see Appendix \ref{appendix:I_general}) that the expression for $F_k^{\text{int}}(\lambda_1,\lambda_2,\chi)$ agrees with the correct limit of equal propagation times; thus,
\begin{align}
\label{Eq36}
\lim_{\chi\rightarrow 0} F^{\text{int}}_{k}(\lambda_1,\lambda_2,\chi) =  \cos\varphi.
\end{align}
Combining the interference and non-interference contributions and setting $\Gamma_1=\Gamma_2$, we get the total tunneling current in the double constriction geometry as $\langle I_T \rangle = 2\langle I_T \rangle_{\Gamma}^{\text{non-int}} + \langle I_T \rangle^{\text{int}}$, whose general expression is given in terms of the $F^{\text{int}}_k$ function as
\begin{align}
\langle I_T \rangle =  - \frac{e^{*2}}{h} V \left( \frac{2\pi |\Gamma|}{v_1^{\alpha/2}v_2^{\beta/2}\hbar} \right)^2 2 (1+F^{\text{int}}_k).
\end{align}
As a caveat, in general, the function $F^{\text{int}}_k$ depends on the potentials $V_1$ and $V_2$, which determine the total charge inside the interferometer. The anyon exchange phase $\varphi$ is contained in $F^{\text{int}}_k$, which takes a simple form  $F^{\text{int}}_k\vert_{\chi=0} = \cos\varphi$ in the case of equal propagation times. As the external magnetic flux changes, the phase jumps due to new anyon excitations. The variation of $F^{\text{int}}_k$ for non-zero values of $\chi$ is shown in Fig.~\ref{fig:4} for the potential bias $V_1=V$ at $V_2=0$. As a reminder, the variable $\lambda = e^* V/\pi k_B \Theta$. 

The structure of Eqs. (\ref{inter-1},\ref{inter-2}) allows identifying two small parameters, which control the validity of the limit of equal propagation times $\chi\rightarrow 0$. First, this is $\chi\sim k_B\Theta(t_2-t_1)/\hbar$, where
$t_{1,2}$ are the travel times between the constrictions along the two channels.
The second parameter originates from the $(\lambda_1-\lambda_2)\tau$ and $\chi\lambda_{1,2}$  terms in the integrals. It is $\chi_V\sim e^*V(t_2-t_1)/\hbar$.
When both $\chi$ and $\chi_V$ are small, the problem allows an exact solution from the next section. The corrections at nonzero $\chi$ and $\chi_V$ can be estimated from the above calculations.

\section{Exact solution by Fermionization}\label{sec:exactsol}

Here we fermionize the problem. This is only possible in two limits: 1) low $\Theta$ and $V_{1,2}$, and 2) equal propagation times between the tunneling contacts along both channels. Case 1) reduces to case 2) with zero propagation times along both channels since in that case the propagation times are much shorter than $\hbar/(e^*V)$ and $\hbar/(k_B\Theta)$.

We start with the action $\mathcal{A}$ given as
\begin{align}
\mathcal{A} = \int dx dt \mathcal{L}_e - \int dt \left( T_1 + T_1^{\dagger} + T_2 + T_2^{\dagger} \right),
\end{align}
where $T_{1,2}$ describe tunneling at the two constrictions and $\mathcal{L}_e$ is the Lagrangian density of the free fields on the right-moving edge,
\begin{align}
\mathcal{L}_e = -\frac{\hbar}{4\pi}\sum_{i=1,2} \left[ \frac{1}{\nu_i} \partial_t \phi_i \partial_x \phi_i + \frac{v_i}{\nu_i} \left(\partial_x\phi_i\right)^2   \right],
\end{align}
where the effective filling factors for the two modes are  $\nu_1 = 1/[(2k-1)(2k+1)]$ and $\nu_2 = 1/[(2k+1)(2k+3)]$, and $k=1,\dots,N-1$. 
The tunneling operators $T_{1,2}$ are defined as
\begin{align}
T_1 &= \Gamma_1  \left(\frac{L}{2\pi}\right)^{-1} :e^{i(2k-1)\phi_1(0)}: :e^{-i(2k+3)\phi_2(0)}: ,\\
T_2 &= \Gamma_2 \left(\frac{L}{2\pi}\right)^{-1} :e^{i(2k-1)\phi_1(l+a)}: :e^{-i(2k+3)\phi_2(l)}:,
\end{align}
where $\Gamma_i$ are the tunneling amplitudes that have units of $[\Gamma_i] =$ energy$\times$length. We also used the fact that $(\alpha+\beta)/2=1$, which reflects the scaling dimensions of the tunneling operators.
It is legitimate to ignore the Klein factors since they enter in the same combination in both tunneling operators.
The Hamiltonian of the full system is given by
\begin{align}
H = \pi \hbar \sum_{i=1,2} \int dx \frac{v_i}{\nu_i e^2} \rho_i^2  + \left( T_1 + T_1^{\dagger} + T_2 + T_2^{\dagger}  \right),
\end{align}
with the commutation relations $\left[ \phi_i(x),\rho_j(x') \right] = -ie\delta_{ij}\nu_i\delta(x-x')$. The tunneling current is defined as the time derivative of the half-difference of the charges of the two modes, 
\begin{align}
I &\equiv \frac{d}{dt} \frac{1}{2} \left( Q_1 - Q_2 \right) \nonumber \\
& = -\frac{i}{\hbar} \frac{e}{2k+1} \left( T_1^{\dagger}-T_1+T_2^{\dagger}-T_2 \right).
\end{align}
In the interaction picture, the tunneling amplitudes get a time dependence as $\Gamma_i\rightarrow \Gamma_i \exp[-ieVt/(2k+1)\hbar]$.

\subsection{Rescaling}\label{sec:exactsol_rescaling}

Next, we rescale the coordinates of the two modes independently. For the field $\phi_i$, we rescale $x\rightarrow \zeta_i x$. The Lagrangian density scales as $\mathcal{L}_e \rightarrow \tilde{\mathcal{L}}_e$, where
\begin{align}
    \tilde{\mathcal{L}}_e = -\frac{\hbar}{4\pi} \sum_{i=1}^2
    \left[ \frac{1}{\nu_i} \partial_t \tilde \phi_i \partial_x \tilde \phi_i + \frac{\tilde v_i}{\nu_i} \left(\partial_x\tilde\phi_i\right)^2 \right],
\end{align}
the rescaled velocities are $\tilde v_i = \zeta_i^{-1}v_i$, and the fields rescale as $\phi_i(x)\rightarrow \tilde \phi_i(x) = \phi_i(\zeta_i x)$. 
The length of the system also rescales.
We will choose the rescaling parameters $\zeta_i$ such that $\tilde{v}_1=\tilde{v}_2$,
which is achieved by choosing $\zeta_i = v_i/u$, where $u$ has the units of speed.
The rescaled action becomes
\begin{align}
    \mathcal{A} &= \int dt dx \tilde{\mathcal{L}}_e - \sum_{i=1}^{2} \int dt \left( \tilde{T}_i + \tilde{T}_i^{\dagger} \right),
\end{align}
where the two co-propagating fields in the edge Lagrangian $\tilde{\phi}_i$ propagate with the same speeds $u$.
Without loss of generality, we will set $u=v_2$. This way we do not need to renormalize the distance $l$ between the constrictions along the second channel. 
The distance $l+a$ along the first channel has to be renormalized,
$l+a\rightarrow l+\tilde a$.
The tunneling operators transform as $T_i\rightarrow \tilde{T}_i$, where
\begin{align}
    \tilde{T}_1(t) &=  \frac{\tilde{\Gamma}_1} {L}  :e^{i(2k-1)\tilde{\phi}_1(0,t)}: :e^{-i(2k+3)\tilde{\phi}_2(0,t)}:
\end{align}
is the tunneling operator at the first constriction. Similarly, the tunneling operator at the second constriction,
\begin{align}
    \tilde{T}_2(t)  &= \frac{\tilde{\Gamma}_2} {L} :e^{i(2k-1)\tilde{\phi}_1(l+a,t)}: :e^{-i(2k+3)\tilde{\phi}_2(l,t)}:,
\end{align}
where the tunneling amplitudes get renormalized to $\tilde{\Gamma}_i = 2\pi\Gamma_i u ~v_1^{-\alpha/2}v_{2}^{-\beta/2}$.
In the interaction representation, charge accumulation in the device results in the phase factor $e^{i\theta}$ (\ref{time-factor}) in $\tilde\Gamma_2$,
where 
\begin{equation}
\label{theta-47}
\theta=\frac{e^*}{\hbar}(V_1t_1-V_2t_2)
\end{equation}
and $t_{1,2}$ are the travel times between the constrictions along the two edges.

The correlation functions of the rescaled fields $\tilde{\phi}_i$ %
are
\begin{align}
    \langle& \tilde{\phi}_i(x,t)\tilde{\phi}_j(0,0) \rangle = -\nu_i\delta_{ij}\nonumber \\
    &~~~\times \text{ln}\left[ \frac{2\pi}{L} \frac{u \hbar}{\pi k_B \Theta}\sin\left( \frac{\pi k_B \Theta}{\hbar} [\epsilon + i (t-x/u)] \right) \right].
\end{align}

What we achieved by rescaling the two fields is that now the fields have the same velocities; however, due to rescaling, the tunneling amplitudes are renormalized as well. We now compute the correlations of the tunneling operators. The auto-correlations are straightforward,
\begin{equation}
    \langle \tilde{T}^{\dagger}_i(t) \tilde{T}^{}_i(0) \rangle  = \frac{|\tilde{\Gamma}_i|^{2} \left( \frac{ k_B \Theta}{2u\hbar} \right)^{2} }{\sin^{2}\left[ \frac{\pi k_B \Theta}{\hbar} (\epsilon + i t) \right]},
\end{equation}
which is the same as for the free-fermion tunneling in the problem of non-interacting fermions.
This suggests that as long as there is only one constriction, the entire problem can be mapped onto a free fermion tunneling problem, with tunneling amplitudes renormalized and the tunneling charge being the fractional quasiparticle charge, since all the algebraic quantities, including the correlators, are the same as for the free fermions. This is not exactly the case when we add a second constriction. We see this from the cross-correlations of the tunneling operators at different constrictions,
\begin{align}
    &\langle \tilde{T}^{\dagger}_{1}(t)\tilde{T}^{}_{2}(0) \rangle = \tilde{\Gamma}_1^{*} \tilde{\Gamma}_2  \left( \frac{ k_B \Theta}{2u\hbar} \right)^{2} \nonumber \\
    &~~~~~~~~~~~~~~~\times \sin^{-\alpha}\left( \frac{\pi k_B \Theta}{\hbar} (\epsilon + i [t-(l+{\tilde a})/{\tilde v}_1]) \right)\nonumber \\
    &~~~~~~~~~~~~~~~\times \sin^{-\beta}\left( \frac{\pi k_B \Theta}{\hbar} [\epsilon + i (t-l/{\tilde v}_2)] \right),
\end{align}
which in general is not the same as in the fermion tunneling case unless 
we consider the limit of equal propagation times, {\it i.e.}, $(l+{\tilde a})/{\tilde v}_1=l/{\tilde v}_2$.
Therefore, in the limit of equal propagation times, the two-point correlation function becomes
\begin{align}
    \langle \tilde{T}^{\dagger}_{1}(t)\tilde{T}^{}_{2}(0) \rangle\Big|_{a=0} &= \tilde{\Gamma}_1^{*} \tilde{\Gamma}_2  \left( \frac{ k_B \Theta}{2u\hbar} \right)^{2} \nonumber \\
    & \times \sin^{-2}\left( \frac{\pi k_B \Theta}{\hbar} [\epsilon+ i (t-l/u)] \right),
\end{align}
which matches with the free-fermion tunneling correlations. %

Remarkably, the time-dependent factor in the tunneling amplitudes $\exp(-ie^*Vt/\hbar)$ is the same as in the interaction representation for the problem of free fermions of charge $e^*$. The charge-accumulation phase (\ref{theta-47})
is also the same as for fermions of charge $e^*$ in the limit of equal propagation times $t_1=t_2$. Thus, to map our problem onto a problem of non-interacting fermions
in the limit $t_1=t_2$, it is sufficient to perform the mapping in the absence of a voltage bias. This will be accomplished in the next subsection with refermionization.

\subsection{Fermionization}\label{sec:exactsol_fermionize}

We first define two new fields $\xi_i$ that are related to the rescaled fields $\tilde\phi_i$ via a change of basis, 
\begin{align}\label{eq:basis_change}
    \begin{pmatrix}
        \tilde\phi_1 \\ \tilde\phi_2 
    \end{pmatrix} = \frac{1}{2(2k+1)} M_k \begin{pmatrix}
        \xi_1 \\ \xi_2
    \end{pmatrix},
\end{align}
where 
the transformation $M_k$ is
\begin{align}
    M_k = \begin{pmatrix}
1+\frac{\sqrt{2k+3}}{\sqrt{2k-1}}& -1+\frac{\sqrt{2k+3}}{\sqrt{2k-1}} \\
-1+\frac{\sqrt{2k-1}}{\sqrt{2k+3}} & 1+\frac{\sqrt{2k-1}}{\sqrt{2k+3}}
\end{pmatrix}.
\end{align}
This corresponds to the commutation relations $[\xi_i(x),\xi_j(x')]=i\pi \delta_{ij}\sgn{(x-x')}$ for the new fields. Thus, the above transformation ensures that the quadratic edge action is mapped onto the bosonized action for fermions at the filling factor $\nu=2$. Next, we define fermion operators
\begin{align}
    \psi_i \equiv \sqrt{\frac{1}{ L}} :\exp(i\xi_i):,
\end{align}
which can be used to recast the action into the form 
\begin{align}
    \mathcal{A} &= i\hbar \int dt dx \sum_{i=1}^2 \psi^{\dagger}_i\left( \partial_t + u\partial_x \right)\psi_i \nonumber \\
    & ~~~ - \int dt \left[ \tilde{\Gamma}_1 \psi^{\dagger}_2(0)\psi_1(0) + \tilde{\Gamma}_2 \psi^{\dagger}_2(l)\psi_1(l) + \text{H.c.} \right].
\end{align}
This action corresponds to the right-moving edge of a $\nu=2$ state of electrons with electron tunneling between the two modes. Hence, in this model, the current operator $I_{\text{F}}$, defined as the time derivative of the half-difference between the charges of the two modes, is $(2k+1)$ times the current in the original model. Thus, $I_F = (2k+1)I$. Also, since in the free-fermion model the commutator of the tunneling operator  and the charge density is $[ Q_1, \exp(i\xi_1(0)-i\xi_2(0))]=-e\exp(i\xi_1(0)-i\xi_2(0))$, in the interaction picture, the tunneling amplitudes get a time dependence of $\tilde{\Gamma}_{i}\rightarrow \tilde{\Gamma}_i \exp(-eVt/\hbar)$ instead of the phase containing $e^*=e/(2k+1)$ in  the original model. To account for this, one needs to divide the voltage bias by $(2k+1)$.

\subsection{Tunneling current}\label{sec:exactsol_electric}
In the fermion action found above, let us first focus on the single constriction case. Thus we set $\Gamma_2 = 0$. Also, to avoid a delta function in the equations of motion, we introduce tunneling in a small vicinity $\epsilon$ of the tunneling contact,
\begin{align}
    \mathcal{A} &= i\hbar \int dt dx \sum_{i=1}^2 \psi^{\dagger}_i\left( \partial_t + u\partial_x \right)\psi_i \nonumber \\
    & - \int dt dx ~\frac{1}{\epsilon}\theta(x(\epsilon -x)) \left[ \tilde{\Gamma}_1 \psi^{\dagger}_2(x)\psi_1(x) + \text{H.c.} \right].
\end{align}
From here, we first find the equations of motion. Since derivatives of $\psi^{\dagger}_{i}$ do not appear, we simply have,
\begin{align}
    i\hbar \left( \partial_t  +  u\partial_x   \right) \psi_1 - \frac{\tilde{\Gamma}_1^*}{\epsilon} \theta(x(\epsilon-x))\psi_2 =0, \\
    i\hbar \left( \partial_t  +  u\partial_x   \right) \psi_2 - \frac{\tilde{\Gamma}_1}{\epsilon} \theta(x(\epsilon-x))\psi_1 =0.
\end{align}
This gives the equation of motion for the modes, $\psi_i(x,t)=\psi_i(x)\exp(-iEt/\hbar)$, where $E=\hbar u q $, 
\begin{align}
    \left( q+ i\partial_x \right) \psi_1(x) &= \frac{\tilde{\Gamma}_1^*}{u\hbar\epsilon} \theta(x(\epsilon-x))\psi_2(x), \\
  \left( q+ i\partial_x \right) \psi_2(x) &= \frac{\tilde{\Gamma}_1}{u\hbar \epsilon} \theta(x(\epsilon-x))\psi_1(x) .
\end{align}
We now have to find solutions to these coupled differential equations. Clearly, there are three regions where we can solve the equations separately. In the region of $x<0$ and $x>\epsilon$, we have a chiral wave equation, and thus the solution is of the form $\psi_i(x)={\text{(const.)}}\times e^{iqx}$. Let us now focus on the region of $0<x<\epsilon$. Here we first decouple the two differential equations as
\begin{align}
    \left( q+i\partial_x\right)^2 \psi_i(x) = \frac{|\tilde\Gamma_1|^2}{(u\hbar\epsilon)^2}\psi_i(x).
\end{align}
Using the ansatz $\psi_i(x)=g_i(x)e^{iqx}$ we obtain the differential equation for $g_i(x)$,
\begin{align}
\partial_x^2g_i(x) = -\frac{|\tilde\Gamma_1|^2}{(u\hbar\epsilon)^2}g_i(x).
\end{align}
Using the boundary conditions, we get two independent solutions in the $\epsilon\rightarrow 0$ limit,
\begin{align}
g_1(x) = 1, ~~~~~~g_2(x) =0;~~~~~~~ &(x<0)\nonumber\\
g_1(x) = t_1, ~~~~~g_2(x) =r_1;~~~~~~ &(x>0)
\end{align}
and 
\begin{align}
g_1(x) = 0, ~~~~~~g_2(x) =1;~~~~~~~ &(x<0)\nonumber\\
g_1(x) = -r_1^*, ~~~~~g_2(x) =t_1;~~~~~~ &(x>0)
\end{align}
where the transmission coefficient $t_1$ and the reflection coefficient $r_1$ \ are given as $t_1 = \cos(|\tilde\Gamma_1|/u\hbar)$ and $r_1 = -ie^{i\bar{\phi}}\sin(|\tilde\Gamma_1|/u\hbar)$, and the phase of the tunneling amplitude $\Gamma_1$ is $\bar{\phi}$. This gives the probability $|r_1|^2$ of tunneling. Note that $t_{1,2}$ was also used for travel times, but no confusion with that notation is possible.

Next, we compute the current. As mentioned earlier, we need to remember that the physical current is $1/(2k+1)$ times the current computed in the free-fermion model, evaluated at the bias $V/(2k+1)$. We use the Landauer-B{\"u}ttiker method to compute the current,
\begin{align}
    I = -|r_1|^2 \frac{1}{(2k+1)}\frac{e}{h} \int_{-\infty}^{\infty} d\varepsilon \left[ \mathcal{F}_1(\varepsilon) - \mathcal{F}_2(\varepsilon) \right],
\end{align}
where $\mathcal{F}_i(\varepsilon)= [\exp((\varepsilon-\mu_i)/k_B \Theta)+1]^{-1}$ is the Fermi-Dirac distribution of the $i$th channel, maintained at the chemical potential $\mu_i$, where $\mu_1=eV_{F}$ and $\mu_2=0$.
Note that the physical voltage bias $V=V_F/(2k+1)$.   
Therefore, the tunneling current in the single constriction case is simply 
\begin{align}
    I = - \frac{e^{*2}}{h} V |r_1|^2,
\end{align} 
which depends on the quasiparticle charge $e^*$. For two constrictions, we only need to know the effective reflection amplitude $r$. Similarly to the calculation for the first constriction, the transmission and reflection amplitudes for the second constriction are given as $t_2 = \cos(|\tilde{\Gamma}_2|/u\hbar)$ and $r_2 = -ie^{i(\bar{\phi}+\varphi)}\sin(|\tilde{\Gamma}_2|/u\hbar)$. We assumed $\Gamma_1 = e^{i{\bar\phi}}|\Gamma_1|$ and $\Gamma_2 = e^{i(\bar{\phi}+\varphi)}|\Gamma_2|$. Then the effective reflection (into drain D2, Fig. \ref{fig:3}) and transmission (into drain D1,  Fig. \ref{fig:3}) amplitudes  are given as $r = t_1 r_2 + r_1 t_2$, and $t = t_1t_2 - r_1^*r_2$. Hence, the current measured in drain D2, which we call the tunneling current, is given as 
\begin{align}
I = -\frac{e^{*2}}{h} V  \left[ |r_1 t_2|^2 + |r_2 t_1|^2 + 2|r_1r_2t_1t_2|\cos\varphi \right],
\end{align}
which is sensitive to the fractional charge $e^*=e/(2k+1)$ because of the Aharonov-Bohm contribution to $\varphi$. The current is sensitive to fractional statistics through the $-4\pi/(2k+1)$ jumps of $\varphi$ with each new quasiparticle inside the interferometer. 

\section{Thermal transport}\label{sec:thermal_transport}

Maintaining the two adjacent modes at different temperatures results in the flow of energy through the quantum point contacts. 
The interference between the heat currents through the two constrictions \cite{therm-int-1,therm-int-2,therm-int-3} depends on 
anyonic statistics in the same way as the interference of charge currents. As introduced in Sec.~\ref{sec:models_setup_thermal}, we assume a zero voltage bias with $V_1=V_2=0$.
The temperatures of the modes $\phi_i$ will be denoted as $\Theta_i$, where $i=1,2$, and the correlation functions of the two modes contain different temperatures. As a warm-up, we first consider transport through a single quantum point contact, so that quasiparticle statistics do not play any role. As in previous sections, we assume that the travel time between the tunneling contacts is shorter for the outer mode. 
Calculations are essentially the same in the opposite case.

\subsection{Single constriction geometry}\label{sec:thermal_transport_single}

For a single constriction, we borrow the setup introduced in Sec.~\ref{sec:models_setup_thermal}, and set $\Gamma_2=0$ and $\Gamma_1=\Gamma\in \mathbb{R}$. We express the tunneling heat current operator (\ref{eq:heat_current_op}) in the interaction picture as
\begin{align}
    J_Q^{(2)}(t) = \Gamma \left( \partial_t \Psi_2^{\dagger}(0,t) \right) \Psi_1(0,t) +  \text{H.c.},
\end{align}
where H.c. denotes the Hermitian conjugate. Using perturbation theory to compute the first non-zero correction to the tunneling heat current, one finds
\begin{align}
    \langle J_Q \rangle^{\text{non-int}}_{\Gamma} &= -\frac{i}{\hbar} \int_{-\infty}^{t} dt' \langle \left[ J_{Q}^{(2)}(t), H_T(t')\right] \rangle_0 .
\end{align}
To compute the average thermal current, we use the following correlation functions,
\begin{align}
    \langle \Psi_{1}^{\dagger}(x,t) \Psi_{1}(0,0) \rangle_0 &= \left[ \mathcal{G}_1^{>}(x,t) \right]^{\alpha}, \\
    \partial_t\langle \Psi_{2}^{\dagger}(x,t) \Psi_{2}(0,0) \rangle_0 &=  \partial_t\left[ \mathcal{G}_2^{>}(x,t) \right]^{\beta},
\end{align}
where the correlation functions of the two modes are defined at different temperatures $\Theta_1$ and $\Theta_2$. Plugging these definitions into the expression for the average thermal tunneling current, we obtain
\begin{align}
    \langle J_Q \rangle^{\text{non-int}}_{\Gamma} &= -\frac{\beta |\Gamma|^2}{\hbar} \frac{(\pi k_B \Theta_2/\hbar)^{2}}{v_{1}^{\alpha}v_{2}^{\beta}} 2n^{\alpha} \nonumber \\
    &\times \int_{-\infty}^{\infty} d\tau  \frac{1}{\sin^{\alpha}\left[ n(\epsilon+i \tau)  \right]}\frac{\cos (  \epsilon+i\tau ) }{\sin^{\beta+1}( \epsilon+i\tau  )} ,
\end{align}
where we redefined the integration variable to $\tau = \pi k_B \Theta_2 (t'-t) /\hbar$, and introduced a variable $n\equiv \Theta_1/\Theta_2$, which controls the amount of heat current flowing through the constriction. We also used $\alpha+\beta=2$ to simplify the integral. We compute the integral in Appendix \ref{appendix:J_special} and obtain

\begin{align}
    \int_{-\infty}^{\infty} d\tau  \frac{1}{\sin^{\alpha}\left[ n(\epsilon+i \tau)  \right]}\frac{\cos (\epsilon+i\tau ) }{\sin^{\beta+1}( \epsilon+i\tau)}  = \frac{\alpha\pi}{6} \left( n^{\beta} - \frac{1}{n^{\alpha}} \right).
\end{align}
Plugging this back into the expression for the tunneling thermal current, we find
\begin{align}
\label{eq:75}
    \langle J_Q \rangle^{\text{n.i.}}_{\Gamma} &= \alpha\beta \frac{\pi |\Gamma|^2}{3\hbar} \frac{(\pi k_B /\hbar)^2}{v_{1}^{\alpha}v_{2}^{\beta}} \left( \Theta_2^2 - \Theta_1^2 \right).
\end{align}
The quadratic dependence on temperature is similar to the free-fermion problem and reflects the scaling dimension of the tunneling operator. Nevertheless, mapping onto a free-fermion problem is no longer possible. The issue is that a change of variables mixes fields at different temperatures and produces new fields whose statistical distribution is not a Gibbs distribution with any temperature.  

We can now extract the tunneling-induced correction to the thermal conductivity $\kappa$ and compare it with the tunneling electrical conductivity $\sigma$ obtained in Section \ref{sec:electric_transport}. We find a Wiedemann-Franz-type relation
\begin{align}
    \frac{\kappa}{\sigma} = (2k-1)(2k+3)~&\frac{\pi^2 k_B^2}{3e^2} \Theta
\end{align}
 with an anomalous Lorenz ratio $L_0 = (2k-1)(2k+3)$. We now focus on the double constriction case, where quasiparticle statistics are essential.

\subsection{Double constriction geometry}\label{sec:thermal_transport_double}

Similar to the electric transport, the total thermal current will now comprise the non-interference and interference contributions. The thermal current operator is
\begin{align}
    J_Q^{(2)}(t) &\equiv \Gamma_1 \left( \partial_t\Psi_2^{\dagger}(0,t) \right) \Psi_1(0,t) + \text{H.c.} \nonumber \\
    &~~~ + \Gamma_2e^{i\varphi} \left( \partial_t\Psi_2^{\dagger}(l,t) \right) \Psi_1(l+a,t) + \text{H.c.}
\end{align}
We introduce the notation $T'_1 \equiv (\partial_t \Psi^{\dagger}_2(0,t)) \Psi_1(0,t)$ and $T'_2 \equiv (\partial_t \Psi^{\dagger}_2(l,t)) \Psi_1(l+a,t)$ for convenience and absorb all phase factors into the definition of the tunneling amplitudes so that the $\Gamma_i$'s are complex. We use perturbation theory to compute the average thermal current. One can express the total tunneling thermal current as
\begin{align}
    \langle J_Q \rangle &= \sum_{i=1}^{2}\langle J_Q \rangle^{\text{non-int}}_{\Gamma_i} + \langle J_Q \rangle^{\text{int}},
\end{align}
where the non-interference contribution is the single constriction result obtained in Section \ref{sec:thermal_transport_single},
\begin{align}
\langle J_Q \rangle^{\text{non-int}}_{\Gamma_i} = \alpha\beta \frac{\pi |\Gamma_i|^2}{3\hbar} \frac{(\pi k_B /\hbar)^2}{v_{1}^{\alpha}v_{2}^{\beta}} \left( \Theta_2^2 - \Theta_1^2 \right).
\end{align} 
What is left is the interference contribution $\langle J_Q \rangle^{\text{int}}$ that will be sensitive to the anyon statistics,
\begin{align}
\langle J_Q \rangle^{\text{int}} & = -\frac{i|\Gamma_1\Gamma_2|}{\hbar} \Bigg[ \int_{-\infty}^{t} dt' \Big( e^{-i\varphi} \langle \left[ T'_1(t), T^{\dagger}_2(t') \right] \rangle_0 - \text{c.c.} \Big) \nonumber\\
&~~~ + \int_{-\infty}^{t} dt' \Big( e^{i\varphi} \langle \left[ T'_2(t),T^{\dagger}_1(t') \right] \rangle_0 - \text{c.c.} \Big) \Bigg],
\end{align}
where c.c. denotes complex conjugate. Notice that $T_{1,2}$ is from the tunneling Hamiltonian $H_T$,
Eq.~(\ref{10-new}), while $T'_{1,2}$ is from the thermal current operator $J_Q$. We substitute into the above equation the two-point correlation function of the tunneling operators given by
\begin{align}
    \langle T'(x,y;t) T^{\dagger}(0,0;0) \rangle & = \left[ \mathcal{G}_1^{>}(x,t) \right]^{\alpha}  \partial_t\left[ \mathcal{G}_2^{>}(y,t) \right]^{\beta},
\end{align}
where Green's functions $\mathcal{G}^{\lessgtr}_i(x,t)$ are defined at their respective temperatures $\Theta_i$, $i=1,2$. Next, we change the integration variable to $\tau \equiv \pi k_B \Theta_2 (t'-t)/\hbar$. In keeping with this definition, we also define the dimensionless propagation time $\tau_1=\pi k_B \Theta_{2}(l+a)/\hbar v_1$ and similarly, $\tau_2=\pi k_B \Theta_2l/\hbar v_2$ for the two modes. Plugging in the correlation functions, we get
\begin{align}
 \langle J_Q \rangle^{\text{int}} & = -\frac{\beta|\Gamma_1\Gamma_2|}{\hbar} \frac{2\cos\varphi}{v_{1}^{\alpha}v_{2}^{\beta} } \left( \frac{\pi k_B \Theta_2}{\hbar} \right)^{2} n^{\alpha}  \\
 &\times \int_{-\infty}^{\infty}d\tau \Bigg[  \frac{\cos[ \epsilon - i (\tau - \chi) ] }{\sin^{\alpha}[n(\epsilon-i\tau)]\sin^{\beta+1}[\epsilon - i (\tau-\chi)]} \nonumber\\
 &~~~~~~~~~~~~+ \frac{\cos[ \epsilon + i (\tau - \chi) ] }{\sin^{\alpha}[n(\epsilon + i\tau)]\sin^{\beta+1}[\epsilon + i (\tau-\chi)]} \Bigg]   \nonumber ,
\end{align}
where we also define the propagation time difference $\chi \equiv \tau_{2}-\tau_{1}$, and the ratio of the temperatures of the two edges as $n\equiv \Theta_1/\Theta_2$. Like before, we need to solve these two integrals with branch points located near $\tau = 0,\chi$. However,  we notice that since $\alpha+\beta=2$, in the limit $\chi\rightarrow 0$, the integrals become straightforward, as we show next. 

\subsubsection{Equal propagation times}\label{sec:thermal_transport_double_equal}
Here we first consider the case when $\chi\rightarrow 0$ that corresponds to the equal propagation times for the two modes. In this limit we get
\begin{align}
    \lim_{\chi\rightarrow 0} \langle J_Q \rangle^{\text{int}} & = - \frac{2\beta|\Gamma_1\Gamma_2|}{\hbar} \frac{2\cos\varphi}{v_{1}^{\alpha}v_{2}^{\beta} } \left( \frac{\pi k_B \Theta_2}{\hbar} \right)^{2} n^{\alpha} \\
    &~~~\times \int_{-\infty}^{\infty} d\tau  \frac{\cos( \epsilon + i \tau  ) }{\sin^{\alpha}[n(\epsilon + i\tau)]\sin^{\beta+1}(\epsilon + i \tau)} .\nonumber
\end{align}
The right-hand side is just the non-interference contribution modulated with the quasiparticle statistics phase, thus, $ \lim_{\chi\rightarrow 0} \langle J_Q \rangle^{\text{int}} = 2\cos\varphi \langle J_Q \rangle^{\text{non-int}}$.
Hence, for identical constrictions with $|\Gamma_1|=|\Gamma_2|=\Gamma$ in the limit of equal propagation times, the total current is simply 
\begin{align}
\label{eq:84}
\lim_{\chi\rightarrow 0} \langle J_Q \rangle = 2(1+\cos\varphi) \langle J_Q \rangle^{\text{non-int}},
\end{align}
which equals the non-interference contribution times an oscillating factor due to localized quasiparticles, as we also obtained in the electrical transport case.

\begin{figure*}
    \centering
    \includegraphics[width=0.8\linewidth]{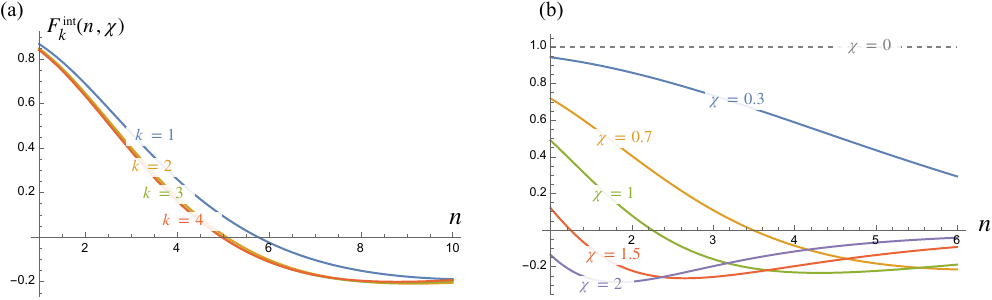}
    \caption{For a topological liquid with a filling factor of $\nu=k/(2k+1)$, the function $F^{\text{int}}_k$ describes the behavior of the thermal current via Eq.~(\ref{Eq85}). We consider $\varphi=0$ here. Panel (a) shows the dependence of $F^{\rm int}_k$ on the temperature ratio $n\equiv \Theta_1/\Theta_2$ at a  fixed difference $\chi=0.5$ of the normalized propagation times between the constrictions along the two channels.  Panel (b) shows the dependence of $F^{\rm int}_2$, which corresponds to a fixed filling factor $\nu=2/5$, on the temperature ratio $n$ at various differences $\chi$ of the travel times along the two paths.  In the limit $\chi\rightarrow 0$,  a simple result (\ref{Eq89}) is recovered. }
    \label{fig:5}
\end{figure*}

\subsubsection{Arbitrary propagation times}
\label{sec:thermal_transport_double_general}
In this section, we compute the interference contribution to the thermal tunneling current with non-zero $\chi$, given $\alpha+\beta=2$. In Appendix \ref{appendix:J_general}, the integrals are solved with these conditions. Using the results from the Appendix, we obtain

\begin{align}\label{Eq85}
    \langle J_Q \rangle^{\text{int}} &= \beta \frac{|\Gamma_1\Gamma_2|}{\hbar} \frac{(\pi k_B \Theta_2/\hbar)^2}{v_1^{\alpha}v_2^{\beta}} \frac{2\pi}{3}\frac{(2k-1)}{(2k+1)}\left( \frac{\Theta_2^2-\Theta_1^2}{\Theta_2^2}\right) \nonumber \\
    & ~~~\times F_{k}^{\text{int}}\left( n, \chi \right),
\end{align}
where $n\equiv \Theta_1/\Theta_2$ and $\chi \equiv \left[l/v_2 - (l+a)/v_1\right] \pi k_B \Theta_2/\hbar$, and the function $F_k^{\text{int}}(n,\chi)$ is defined as
\begin{widetext}
    \begin{align}
        F_k^{\text{int}}(n,\chi)&=-6\frac{n^{\alpha}}{ \alpha\pi(1-n^2)}\sin\left(\frac{2\pi}{2k+1}\right) \cos\varphi \Bigg[ \frac{1}{\beta\chi^{\beta}\sinh^{\alpha}(n\chi)} + \frac{n\alpha \cosh(n\chi)}{(\beta-1)\sinh^{\alpha+1}(n\chi)\chi^{\beta-1}} \nonumber \\
& ~~~+ \int_{0}^{\chi}d\tau \left( \frac{1}{(\chi-\tau)^{\beta+1}\sinh^{\alpha}(n\chi)} + \frac{n\alpha \cosh(n\chi)}{(\chi-\tau)^{\beta}\sinh^{\alpha+1}(n\chi)} - \frac{\cosh(\chi-\tau)}{\sinh^{\alpha}(n\tau)\sinh^{\beta+1}(\chi-\tau)} \right) \Bigg] ,
    \end{align}
\end{widetext}
and $k=1,\dots,N-1$ with the bulk filling factor $\nu_b=N/(2N+1)$. As discussed in the Appendix, the limit of equal propagation times works out, {\it i.e.},
\begin{align}
\lim_{\chi\rightarrow 0} F^{\text{int}}_{k}(n,\chi) = \cos\varphi .%
\end{align}
Fig.~\ref{fig:5} shows the variation of the function $F^{\text{int}}_k(n,\chi)$  with respect to the temperature ratio $n\equiv \Theta_1/\Theta_2$ at different non-zero values of $\chi$. Combining the interference and non-interference contributions, we get the total tunneling current in the double constriction geometry (for identical constrictions such that $\Gamma_i=\Gamma$) as
\begin{align}
\langle J_Q \rangle &= 2\langle J_Q \rangle^{\text{non-int}}_{\Gamma} + \langle J_Q \rangle^{\text{int}},\\
&= \alpha\beta \frac{\pi |\Gamma|^2}{3\hbar} \frac{(\pi k_B /\hbar)^2}{v_{1}^{\alpha}v_{2}^{\beta}} \left( \Theta_2^2 - \Theta_1^2 \right) 2 (1+ F^{\text{int}}_k), \nonumber
\end{align}
which in general is expressed in terms of the function $F^{\text{int}}_k(n,\chi)$.  Making use of the limit $\lim_{\chi\rightarrow 0} F^{\text{int}}_k(n,\chi)$, the limit of equal propagation times for the total thermal current
gives the following for two identical constrictions:
\begin{align}\label{Eq89}
    \lim_{\chi\rightarrow 0} \langle J_Q \rangle &= 2(1+\cos\varphi) \frac{(2k-1)}{(2k+1)}\frac{(2k+3)}{(2k+1)} \nonumber \\
    &~~~~~~~~~ \times \frac{\pi|\Gamma|^2}{3\hbar}\frac{(\pi k_B /\hbar)^2}{v_1^{\alpha} v_2^{\beta}} \left(\Theta_2^2-\Theta_1^2 \right),
\end{align}
where the integer $k=1,\dots,N-1$ labels the pair of the channels used in the device. This agrees with equations~(\ref{eq:75},\ref{eq:84}).

\section{Electrical noise}\label{sec:noise}

In this section, we focus on the electric current noise in the interferometer \cite{review-FH,noise-review}. From the fluctuation-dissipation theorem, we expect that the zero-frequency spectral noise of the system should be related to the average tunneling current. We define noise as the tunneling current-current correlation $S(t) = \langle \{I_T(t),I_T(0)\} \rangle$, and thus the noise spectrum is defined as its Fourier transform, $S(\omega) = \int dt \exp(-i\omega t) S(t)$. We will focus on the limit $\omega\rightarrow 0$. As usual, this limit cannot be obtained by a direct substitution of $\omega=0$ since such a substitution results in a divergent integral in the case of a nonzero average current.

\subsection{Perturbation theory}

We start with the lowest nonzero contribution in powers of the tunneling amplitudes $\Gamma_{1,2}$.
The electric current can be expressed as 
\begin{equation}
I=e^*(R_{2\rightarrow 1}-R_{1\rightarrow 2}),
\end{equation}
where $R_{i\rightarrow j}$ are the rates of quasiparticle tunneling from channel $i$ to channel $j$. According to the detailed balance principle,
\begin{equation}
R_{1\rightarrow 2}=\exp\left(\frac{e^*V}{k_B\Theta}\right) R_{2\rightarrow 1}.
\end{equation}
Expressing the noise in terms of the matrix elements of the current operators between energy eigenstates, we discover that \cite{levitov2004}
\begin{equation}
\label{noise-1}
S(\omega\rightarrow 0)=2(e^*)^2(R_{1\rightarrow 2}+R_{2\rightarrow 1})=
2e^*|I|\coth\left(\frac{e^*V}{2k_B\Theta}\right).
\end{equation}

We have computed the noise of the tunneling current. Normally, current is measured
in one of the drains. The noise $S_i(\omega\rightarrow 0)$ in the drain in channel $i$ can be expressed from the above result (\ref{noise-1}) with the help of a nonequilibrium fluctuation-dissipation theorem \cite{fdt1,fdt2,fdt3,safi2011:PhysRevB.84.205129}, which states that
\begin{align}
\label{FDT}
    S_i(\omega\rightarrow 0) = S + 4k_B \Theta \frac{\partial \langle I_T \rangle}{\partial V} + 4G_i k_B \Theta,
\end{align}
where $G_{i}= \nu_i e^2/h$ and the result for the current can be found in Section \ref{sec:electric_transport}.

\subsection{Exact solution for electrical noise}\label{sec:noise_exactsol}

As demonstrated in Section \ref{sec:exactsol}, the electric transport problem can be mapped onto the free-fermion tunneling problem in the limit of equal propagation times. 
At sufficiently low voltages and temperatures, the limit of equal propagation times always applies since the travel times between the constrictions along the two channels
become much smaller than $\hbar/(e^*V)$ and $\hbar/(k_B\Theta)$.

The mapping allows us to find exact solutions for the average current and noise. In this section, we find the exact expression for noise in our Mach-Zehnder setup. In Section \ref{sec:exactsol}, we computed an effective tunneling rate between the channels, $|r|^2=|r_2t_1+r_1t_2|^2$, where the expressions for the amplitudes $r_i$ and $t_i$ in terms of the tunneling amplitudes $\Gamma_i$ can be found in Section \ref{sec:exactsol_electric}. 

The zero-frequency noise measures the current correlations in one of the drains, D1 or D2. It is defined as
\begin{align}
    S_i(\omega\rightarrow 0) = \int_{-\infty}^{\infty} dt \langle \{ \Delta I_{D_i}(t),\Delta I_{D_i}(0) \} \rangle,
\end{align}
where $\Delta I_{D_i}(t) \equiv I_{D_i} - \langle I_{D_i} \rangle$, and $I_{D_i}$ is the current measured in the $i$th drain ($i=1,2$). 
As discussed above, a fluctuation-dissipation theorem (\ref{FDT}) holds in chiral systems.
The noise $S$ of the tunneling current is defined in terms of the current correlations,
\begin{align}
    S(\omega\rightarrow 0) = \int_{-\infty}^{\infty} dt \langle \{ \Delta I_{T}(t),\Delta I_{T}(0) \} \rangle.
\end{align}
Once the problem is mapped onto the free-fermion tunneling problem, the noise $S$ in the original setup is found by evaluating the noise of the tunneling current in the free-fermion case at $V/(2k+1)$. We also need to remember that the current operators in the free-fermion and original models are related as $I_F = (2k+1)I$. 

The noise $S_{T,f}$ of the tunneling current in a free-fermion problem \cite{martin1992,feldman:2017} has been computed as a function of the tunneling probability $|r|^2$:

\begin{align}
    S_{T,f}&(\omega\rightarrow 0)= 4G_0k_B \Theta |r|^2 \\
    &~+ 2eVG_0|r|^2(1-|r|^2) \left[ \coth\left( \frac{eV}{2k_B \Theta} - \frac{2k_B \Theta}{eV} \right) \right]. \nonumber
\end{align}
The tunneling noise in the original model is therefore given by $S_T = S_{T,f}(V/(2k+1))/(2k+1)^2$. We computed the tunneling current in section \ref{sec:exactsol_electric}. Therefore, one obtains from the fluctuation-dissipation theorem (\ref{FDT}) that
\begin{align}\label{eq:noise}
    S_i&(\omega\rightarrow 0) = 4\nu_i G_0 k_B \Theta \\
    &~ + \frac{2e^* V G_0}{(2k+1)^2}|r|^2(1-|r|^2)  \left[ \coth\left( \frac{e^*V}{2k_B \Theta} - \frac{2k_B \Theta}{e^*V} \right) \right]. \nonumber 
\end{align}
The first term is the Johnson-Nyquist noise, which depends on the conductance of the channel $G=\nu_iG_0$. The second term is the shot noise due to tunneling between the two copropagating channels and depends on $e^*$. 

The shot noise is sensitive to the quasiparticle charge $e^*=e/(2k+1)$ and quasiparticle statistics via two non-zero harmonics in its dependence on the sum of the statistical and Aharonov-Bohm phases:
\begin{align}
    |r|^2(1-|r|^2) &= R_0(1-R_0)-2R_1^2 \nonumber \\
    &~~+ 2R_1(1-2R_0)\cos\varphi - 2R_1^2 \cos 2\varphi,
\end{align}
where the coefficients $R_0 \equiv |r_1t_2|^2+|r_2t_1|^2$ and $R_1\equiv |r_1r_2t_1t_2|$.
We emphasize that expression (\ref{eq:noise}) is an exact result for the electrical noise in our setup. 

\section{Conclusions and Discussion}\label{sec:conclusions}

Starting with Newton's rings \cite{Micrographia} in the 1660s, many types of optical interferometers have been developed. A Fabry-P{\'e}rot device \cite{fabry1899} allows light to travel distances much longer than the interferometer size due to multiple reflections from two mirrors. A Mach-Zehnder interferometer \cite{zehnder1891,mach1892} spatially separates the interference beams so that they can be individually manipulated. An electronic version of a Fabry-P{\'e}rot setup \cite{chamon1997:PhysRevB.55.2331} involves multiple interfering trajectories made of many loops around the device in a close resemblance to an optical device. This creates a challenge for probing anyons since the statistical phase, accumulated by anyons on a path around the interferometer, depends on the number of loops. A standard anyonic Mach-Zehnder setup \cite{MZ-heiblum} is free from this challenge, but such a device is harder to fabricate because of the need to place a drain in its middle. The topological charge in the device changes after each tunneling event. This results in rich and interesting physics but leads to a nontrivial and sometimes complicated relation between fractional statistics and observables \cite{review-FH,ma2016-16,zucker2016}. 

In this work, we address a setup that closely parallels optical Mach-Zehnder interferometry and does not involve a drain inside the device. This makes the interpretation of the transport data very straightforward. As an added advantage, 
the area of the device is less sensitive to the number of trapped anyons than in the standard Fabry-P{\'e}rot technique. Moreover, in a nice surprise, an exact solution for electric current and noise can be obtained under realistic conditions.
The solution is possible if the travel times between the two tunneling contacts are equal on both branches of the interferometer. It also holds in the limit of low temperature and voltage when the interferometer is much shorter than the length scale $\hbar v/{\rm max}(eV, k_B \Theta)$ set by the voltage $V$ and the temperature $\Theta$, since in the latter case the travel time can be neglected. Of course, the exactly solvable model contains only the most relevant quadratic terms in the edge Hamiltonian and the most relevant tunneling operators at the constrictions. This is justified in the low-energy limit.
Beyond that limit, interferometry is of limited utility since the phases accumulated by anyons in the device depend on their energy. As a result, the interferometry picture is smeared. 

To probe the statistics of anyons, one needs either to change the magnetic field through the device or modify the area of the interferometer with the help of side gates. In both cases, new anyons enter between the interfering channels from time to time. Every time, the interference phase jumps by $-4\pi/(2k+1)$. Remarkably, the jump can be easily read out from the comparison of the data with exact results for the current and noise even for strong inter-channel tunneling when the visibility of the interference signal is high.

We used perturbation theory to address interferometry beyond the equal-time propagation limit. The result is quantitatively valid for weak tunneling between the interferometer channels. It also helps to estimate corrections to the exact solution for strong tunneling when the propagation times are different. 

Fractional statistics affect all transport properties of an interferometer, including energy transport, and it is instructive to compare electric and heat transport through the same device. For that reason, we computed the heat current through the interferometer subjected to a temperature gradient. An exact solution is no longer possible because the refermionized modes do not follow Gibbs distributions. This highly non-equilibrium situation resembles thermal transport in junctions of Majorana channels on the surface of a topological insulator \cite{shapiro2017}.

We have focused on the Jain states. It is straightforward to generalize our calculations to other Abelian states. A more challenging question involves non-Abelian statistics. Many advantages of the new setup apply equally in the Abelian and non-Abelian cases. At the same time, it is presently unclear if an exact solution remains available.

\section*{Acknowledgments}

We thank M. Heiblum for useful discussions. The research by NB, ZW, and DEF was supported in part by the National Science Foundation under grant No. DMR-2204635. SV was supported in part by National Science Foundation through Grant No. DMR-2004825.\\

The data underlying Fig. \ref{fig:4} and Fig. \ref{fig:5} in this paper are available at Zenodo \cite{batra_2025_data}.

\appendix
\onecolumngrid

\section{Integrals for electric current at equal propagation times}\label{appendix:I_special}

\begin{figure*}
    \centering
    \includegraphics[width=0.75\linewidth]{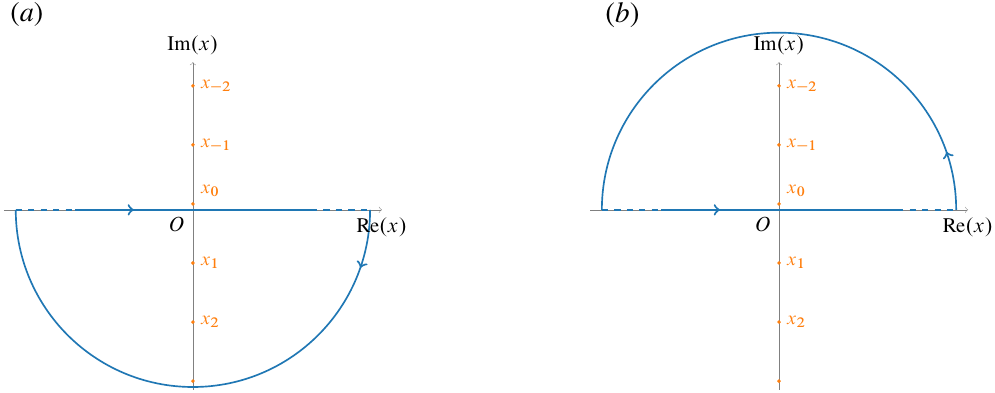}
    \caption{Integration contours (a) for $\mathbb{I}_1$ and (b) $\mathbb{I}_2$.}
    \label{fig:contour-12}
\end{figure*}

In the main text, we define the Fourier transforms of Green's functions $\tilde{G}^{>}(\omega)=\tilde{G}^{<}(-\omega)=\int d\eta \exp(-i\omega t) \langle T(t)T^{\dagger}(0) \rangle_0$. The two-point correlation function of the tunneling operators is defined as
\begin{align}
    \langle T(t)T^{\dagger}(0) \rangle_0 = \left(\frac{\pi k_B \Theta}{\hbar}\right)^2 v_1^{-\alpha}v_2^{-\beta} \sin^{-2}\left( \frac{\pi k_B \Theta}{\hbar} \left( \epsilon + i t \right) \right),
\end{align}
where the rational variables $\alpha$ and $\beta$ satisfy $\alpha + \beta =2$. We now compute the Fourier transform $\tilde G^>(\omega)$,
\begin{align}
\tilde G^>(\omega) 
&= \int_{-\infty}^{\infty} dt e^{-i\omega t} \langle T(0,0;t)T^{\dagger}(0,0;0) \rangle 
= \frac{(\pi k_B \Theta/\hbar)^2}{v_{1}^{\alpha}v_{2}^{\beta}}  \int_{-\infty}^{\infty} dt e^{-i\omega t} \sin^{-2} \left( \frac{\pi k_B \Theta}{\hbar} (\epsilon + i t) \right) \nonumber\\
&= \frac{(\pi k_B \Theta/\hbar)}{v_{1}^{\alpha}v_{2}^{\beta}} \int_{-\infty}^{\infty} dx e^{-i\lambda x} \sin^{-2}(\epsilon+ix) 
\equiv \frac{(\pi k_B \Theta/\hbar)}{v_{1}^{\alpha}v_{2}^{\beta}} \times  \mathbb{I}_1,
\end{align}
where we change the integration variable and define $\lambda = \hbar \omega/\pi k_B \Theta$. Since the exponents $\alpha$ and $\beta$ add up to an integer, we can find an analytical solution to the integral straightforwardly. Similarly, for $\tilde{G}^{<}(\omega)$, we have
\begin{align}
\tilde G^<(\omega) = \tilde G^>(-\omega) 
&= \int_{-\infty}^{\infty} dt e^{i\omega t} \langle T(0,0;t)T^{\dagger}(0,0;0) \rangle 
= \frac{(\pi k_B \Theta/\hbar)^2}{v_{1}^{\alpha}v_{2}^{\beta}}  \int_{-\infty}^{\infty} dt e^{i\omega t} \sin^{-2} \left( \frac{\pi k_B \Theta}{\hbar} (\epsilon + i t) \right)\nonumber \\
&= \frac{(\pi k_B \Theta/\hbar)}{v_{1}^{\alpha}v_{2}^{\beta}} \int_{-\infty}^{\infty} dx e^{i\lambda x} \sin^{-2}(\epsilon+ix) \equiv \frac{(\pi k_B \Theta/\hbar)}{v_{1}^{\alpha}v_{2}^{\beta}} \times  \mathbb{I}_2.
\end{align}
Note that since $\omega_q= e^*V/\hbar>0$, we need to compute the above two integrals separately assuming $\lambda>0$. Let us first compute $\mathbb{I}_1= \int dx e^{-i\lambda x} \sin^{-2}(\epsilon+ix) $ by analytically continuing it to the complex plane. In the complex plane, the poles of the integrand are located at $x_n = i(\epsilon - n\pi)$ with $n\in \mathbb{Z}$. Assuming $\lambda>0$, we close the integration contour (Fig.~\ref{fig:contour-12}) in the lower half-plane enclosing the poles corresponding to $n=\{1,2,\dots\}$. 

We find that
\begin{align}
\mathbb{I}_1 = \int_{-\infty}^{\infty}dx e^{-i\lambda x} \sin^{-2}(\epsilon+ix) = \oint_{\text{lhp}} dx e^{-i\lambda x} \sin^{-2}(\epsilon+iz) = -2\pi i \sum_{n=1}^{\infty} \text{Res}_n\left( e^{-i\lambda x}\sin^{-2}(i(x-x_n)) ,x_n\right).
\end{align}
Since around $x_n$, we have a second-order pole, the residue at $x_n$ is $\text{Res}_n (f(x),x_n) = i\lambda e^{-i\lambda x_n}$. Plugging this back gives
\begin{align}
\mathbb{I}_1 = 2\pi \sum_{n=1}^{\infty} \lambda e^{-i\lambda x_n} = 2\pi \lambda \sum_{n=1}^{\infty} e^{\lambda \epsilon} e^{-\lambda n \pi} = 2\pi \lambda e^{\lambda \epsilon} \frac{e^{-\lambda \pi}}{1-e^{-\pi \lambda}} \xrightarrow{\epsilon\rightarrow 0} \frac{2\pi \lambda e^{-\lambda \pi}}{1-e^{-\pi \lambda}}.
\end{align}
Similarly, for $\mathbb{I}_2$, we need to close the contour in the upper half-plane (Fig.~\ref{fig:contour-12}), thus enclosing the poles corresponding to $n=\{0,-1,-2,\dots\}$, where again $x_n = i(\epsilon-n\pi)$, and the residue is $\text{Res}_n (f(x),x_n) = -i\lambda e^{i\lambda x_n}$. Therefore we get
\begin{align}
\mathbb{I}_2 = \oint_{\text{uhp}} dx e^{i\lambda x} \sin^{-2}(\epsilon+iz) = 2\pi i \sum_{n=0}^{-\infty} \text{Res}_n\left( e^{i\lambda x}\sin^{-2}(i(x-x_n)) ,x_n\right) = 2\pi\lambda  \sum_{n=0}^{-\infty}  e^{i\lambda x_n} = 2\pi\lambda e^{-\lambda \epsilon} \frac{1}{1-e^{-\lambda \pi}} .
\end{align}
Using these results, we obtain the first non-zero correction to the tunneling current,
\begin{align}
\langle I_T \rangle^{\text{non-int}}_{\Gamma} = \frac{e^* |\Gamma|^2}{\hbar^2} \left[ \tilde G^>(\omega_q) - \tilde G^<(\omega_q)  \right] = - \frac{e^* |\Gamma|^2}{\hbar^2} \frac{(\pi k_B \Theta/\hbar)}{v_{1}^{\alpha}v_{2}^{\beta}} 2\pi \frac{\hbar \omega_q}{\pi k_B \Theta} = - \frac{1}{(2k+1)^2} \frac{e^2}{h} V \left( \frac{2\pi |\Gamma|}{v_1^{\alpha/2}v_2^{\beta/2}\hbar} \right)^2,
\end{align}
where the exponents $\alpha = (2k-1)/(2k+1)$ and $\beta = (2k+3)/(2k+1)$, and $k=1,\dots,N-1$ defines the two propagating edge modes of the $\nu_b=N/(2N+1)$ FQH liquid.

\section{Integrals for electric current at different propagation times}\label{appendix:I_general}

\begin{figure}[htbp]
    \centering
    \includegraphics[width=0.5\linewidth]{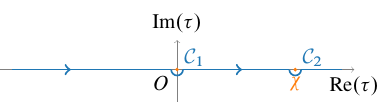}
    \caption{The integration contour for $\mathbb{I}_1$.}
    \label{fig:contour-3}
\end{figure}

\begin{figure}[htbp!]
    \centering
    \includegraphics[width=0.5\linewidth]{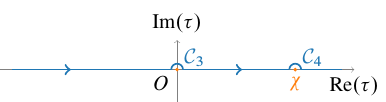}
    \caption{The integration contour for $\mathbb{I}_2$.}
    \label{fig:contour-4}
\end{figure}

We now focus on the general case where the propagation times in the two modes may be different. We need to solve the integrals in the following expression with $\alpha+\beta=2$ and $\alpha,\beta>0$: 
\begin{align}
\langle I_T(t) \rangle^{\text{int}} &= -2i\frac{e^*|\Gamma_1\Gamma_2|}{\hbar^2} \frac{(\pi k_B \Theta/\hbar)}{ v_{1}^{\alpha} v_{2}^{\beta} }  \Bigg[ 
\int_{-\infty}^{\infty}d\tau  \frac{\sin \left(\lambda \tau-\phi'\right)} {\sin^{\alpha}(\epsilon+i\tau) \sin^{\beta}\left[\epsilon+i(\tau-\chi)\right]} 
- \int_{-\infty}^{\infty}d\tau \frac{ \sin\left(\lambda \tau - \phi'\right)}{\sin^{\alpha}(\epsilon-i\tau) \sin^{\beta}\left[\epsilon-i(\tau-\chi)\right]} \Bigg].
\end{align}
Notice that since $\alpha= (2k-1)/(2k+1)$ and $\beta=(2k+3)/(2k+1)$, we also have  $\alpha<\beta$, which will be useful. Assuming $\chi\ge 0$, we now proceed to solve the first of the two integrals
\begin{align}
\mathbb{I}_1 & =  \int_{-\infty}^{\infty}d\tau  \frac{\sin \left(\lambda \tau-\phi'\right)} {\sin^{\alpha}(\epsilon+i\tau) \sin^{\beta}\left[\epsilon+i(\tau-\chi)\right]} \nonumber \\
& = \left[ \int_{-\infty}^{-\epsilon}d\tau +  \int^{}_{\mathcal{C}_1}d\tau +  \int^{\chi-\epsilon}_{\epsilon}d\tau +  \int^{}_{\mathcal{C}_2}d\tau +  \int_{\chi+\epsilon}^{\infty}d\tau  \right]  \frac{\sin \left(\lambda \tau-\phi'\right)} {\sin^{\alpha}(i\tau) \sin^{\beta}\left[i(\tau-\chi)\right]},
\end{align}
where $\mathcal{C}_{1,2}$ are semicircle contours (Fig.~\ref{fig:contour-3}) in the lower half-plane around $\tau=0,\chi$. 
We now separate the phase and the absolute value of the integrals for each of the segments $(-\infty,-\epsilon)$, $\mathcal{C}_1$, $(+\epsilon,\chi-\epsilon)$, $\mathcal{C}_2$, and $(\chi+\epsilon,+\infty)$. First,
\begin{align}
\int_{-\infty}^{-\epsilon}d\tau \frac{\sin \left(\lambda \tau-\phi'\right)} {\sin^{\alpha}(i\tau) \sin^{\beta}\left[i(\tau-\chi)\right]} = (-i)^{-\alpha} (-i)^{-\beta}\int_{-\infty}^{-\epsilon}d\tau \frac{\sin(\lambda\tau -\phi')}{\sinh^{\alpha}(-\tau)\sinh^{\beta}(\chi-\tau)} = \int_{\epsilon}^{\infty}d\tau \frac{\sin(\lambda\tau+\phi')}{\sinh^{\alpha}(\tau)\sinh^{\beta}(\tau+\chi)}.
\end{align}
Then for $\mathcal{C}_1$, we change the variable $\tau$, $\tau=\epsilon e^{i\theta}$, where $\theta$ runs from $-\pi\rightarrow 0$,
\begin{align}
\int_{\mathcal{C}_1}d\tau \frac{\sin \left(\lambda \tau-\phi'\right)} {\sin^{\alpha}(i\tau) \sin^{\beta}\left[i(\tau-\chi)\right]} = \int_{-\pi}^{0}d\theta \frac{i\epsilon e^{i\theta}\sin(\lambda \epsilon e^{i\theta}-\phi')}{\sin^{\alpha}(i\epsilon e^{i\theta})\sin^{\beta}(i\epsilon e^{i\theta}-i\chi)} \sim \frac{\sin(\phi')}{\sinh^{\beta}(\chi)}\epsilon^{1-\alpha} \int_{-\pi}^{0}d\theta e^{i(1-\alpha)\theta} + \mathcal{O}(\epsilon^{(2-\alpha)}) \xrightarrow{\epsilon\rightarrow 0}0.
\end{align}
 We made use of the fact that since $\alpha+\beta=2$ and $\beta>\alpha$, so $\alpha<1$, and thus $(1-\alpha)>0$. Then only the first term in the series expansion
\begin{align}
\frac{1}{\sin^{\alpha}(z)\sin^{\beta}(z+w)}\Bigg|_{z=0} = \frac{1}{z^{\alpha}\sin^{\beta}(w)}+ \mathcal{O}(z^{(1-\alpha)})
\end{align}
diverges. Next, we have the integral
\begin{align}
\int_{\epsilon}^{\chi-\epsilon}d\tau \frac{\sin \left(\lambda \tau-\phi'\right)} {\sin^{\alpha}(i\tau) \sin^{\beta}\left[i(\tau-\chi)\right]} = \int_{\epsilon}^{\chi-\epsilon}d\tau \frac{(i)^{-\alpha}(-i)^{-\beta}\sin(\lambda\tau-\phi')}{\sinh^{\alpha}(\tau)\sinh^{\beta}(\chi-\tau)} = \int_{\epsilon}^{\chi-\epsilon}d\tau \frac{e^{i(1-\alpha)\pi}\sin(\lambda\tau-\phi')}{\sinh^{\alpha}(\tau)\sinh^{\beta}(\chi-\tau)}.
\end{align}
Then for $\mathcal{C}_2$, we change the variable $\tau$, $\tau=\chi+\epsilon e^{i\theta}$, where $\theta$ runs from $-\pi\rightarrow 0$,
\begin{align}
\int_{\mathcal{C}_2}d\tau \frac{\sin \left(\lambda \tau-\phi'\right)} {\sin^{\alpha}(i\tau) \sin^{\beta}\left[i(\tau-\chi)\right]} &= \int_{-\pi}^{0}d\tau \frac{i\epsilon e^{i\theta}\sin(\lambda\epsilon e^{i\theta}+\lambda\chi-\phi')}{\sin^{\alpha}(i\epsilon e^{i\theta}+i\chi)\sin^{\beta}(i\epsilon e^{i\theta})} = \frac{\sin(\lambda\chi - \phi')}{\sinh^{\alpha}(\chi)\epsilon^{\beta-1}}e^{-i\frac{\pi}{2}}\int_{-\pi}^{0}d\theta e^{-i(\beta-1)\theta} + \mathcal{O}(\epsilon^{2-\beta})\nonumber\\
&= \frac{\sin(\lambda\chi - \phi')}{\sinh^{\alpha}(\chi)~\epsilon^{\beta-1}}\frac{1}{(\beta-1)}\left[ 1-e^{i(\beta-1)\pi} \right] + \mathcal{O}(\epsilon^{2-\beta}).
\end{align}
From the conditions on $\alpha$ and $\beta$, we have $2-\beta>0$. Hence, the first two terms in the series expansion
\begin{align}
\frac{1}{\sin^{\alpha}(z+w)\sin^{\beta}(z)}\Bigg|_{z=0} = \frac{1}{z^{\beta}\sin^{\alpha}(w)} - \frac{\alpha \cos(w)}{z^{\beta-1}\sin^{1+\alpha}(w)} + \mathcal{O}(z^{2-\beta})
\end{align}
diverge. Finally, in the last interval,
\begin{align}
\int_{\chi+\epsilon}^{\infty} d\tau \frac{\sin \left(\lambda \tau-\phi'\right)} {\sin^{\alpha}(i\tau) \sin^{\beta}\left[i(\tau-\chi)\right]} 
= \int_{\chi+\epsilon}^{\infty} d\tau \frac{ i^{-\alpha}i^{-\beta}  \sin \left(\lambda \tau-\phi'\right)} {\sinh^{\alpha}(\tau) \sinh^{\beta}(\tau-\chi)} 
= - \int_{\chi+\epsilon}^{\infty} d\tau \frac{  \sin \left(\lambda \tau-\phi'\right)} {\sinh^{\alpha}(\tau) \sinh^{\beta}(\tau-\chi)},
\end{align}
where we used $\alpha + \beta =2$. Let us now do the same for the second integral
\begin{align}
\mathbb{I}_2 &= \int_{-\infty}^{\infty}d\tau  \frac{\sin \left(\lambda \tau-\phi'\right)} {\sin^{\alpha}(\epsilon-i\tau) \sin^{\beta}\left[\epsilon-i(\tau-\chi)\right]} \nonumber\\
& = \left[ \int_{-\infty}^{-\epsilon}d\tau +  \int^{}_{\mathcal{C}_3}d\tau +  \int^{\chi-\epsilon}_{\epsilon}d\tau +  \int^{}_{\mathcal{C}_4}d\tau +  \int_{\chi+\epsilon}^{\infty}d\tau  \right]  \frac{\sin \left(\lambda \tau-\phi'\right)} {\sin^{\alpha}(-i\tau) \sin^{\beta}\left[-i(\tau-\chi)\right]},
\end{align}
where now $\mathcal{C}_{3,4}$ are semicircle contours (Fig.~\ref{fig:contour-4}) in the upper half-plane around $\tau=0,\chi$. 
First, we have the contribution
\begin{align}
\int_{-\infty}^{-\epsilon}d\tau \frac{\sin \left(\lambda \tau-\phi'\right)} {\sin^{\alpha}(-i\tau) \sin^{\beta}\left[-i(\tau-\chi)\right]} 
= \int_{-\infty}^{-\epsilon}d\tau \frac{(-i)^{-\alpha}(-i)^{-\beta}\sin(\lambda \tau - \phi')}{\sinh^{\alpha}(-\tau)\sinh^{\beta}(\chi-\tau)} 
= \int^{\infty}_{\epsilon}d\tau \frac{\sin(\lambda \tau + \phi')}{\sinh^{\alpha}(\tau)\sinh^{\beta}(\chi+\tau)}.
\end{align}
Then for the contour $\mathcal{C}_3$ we change the variable $\tau=\epsilon e^{i\theta}$, where $\theta$ runs from $\pi\rightarrow 0$,
\begin{align}
\int_{\mathcal{C}_3}d\tau \frac{\sin \left(\lambda \tau-\phi'\right)} {\sin^{\alpha}(-i\tau) \sin^{\beta}\left[-i(\tau-\chi)\right]} = \int_{\pi}^{0}d\theta \frac{i\epsilon e^{i\theta} \sin(\lambda \epsilon e^{i\theta}-\phi') }{\sinh^{\alpha}(-i\epsilon e^{i\theta})\sin^{\beta}(-i\epsilon e^{i\theta}+i\chi)}  \sim \frac{\sin(\phi')}{\sinh^{\beta}(\chi)}\epsilon^{1-\alpha} \int_{\pi}^{0}d\theta e^{i(1-\alpha)\theta} + \mathcal{O}(\epsilon^{2-\alpha}) \xrightarrow{\epsilon\rightarrow 0} 0,
\end{align}
where again we made use of a series expansion as before. Next, we have the contribution
\begin{align}
\int^{\chi-\epsilon}_{\epsilon}d\tau \frac{\sin \left(\lambda \tau-\phi'\right)} {\sin^{\alpha}(-i\tau) \sin^{\beta}\left[-i(\tau-\chi)\right]} 
= \int_{\epsilon}^{\chi-\epsilon}d\tau \frac{ (-i)^{-\alpha}i^{-\beta}\sin(\lambda\tau-\phi')}{\sinh^{\alpha}(\tau)\sinh^{\beta}(\chi-\tau)}
= \int_{\epsilon}^{\chi-\epsilon}d\tau \frac{ e^{-i(1-\alpha)\pi}\sin(\lambda\tau-\phi')}{\sinh^{\alpha}(\tau)\sinh^{\beta}(\chi-\tau)}.
\end{align}
Again, for the contour $\mathcal{C}_4$, we change the integration variable $\tau=\chi+\epsilon e^{i\theta}$, where $\theta$ runs from $\pi\rightarrow 0$,
\begin{align}
\int_{\mathcal{C}_4}d\tau \frac{\sin \left(\lambda \tau-\phi'\right)} {\sin^{\alpha}(-i\tau) \sin^{\beta}\left[-i(\tau-\chi)\right]} &= \int_{\pi}^{0}d\theta \frac{i\epsilon e^{i\theta} \sin(\lambda \epsilon e^{i\theta}+\lambda\chi - \phi')}{\sin^{\alpha}(-i\epsilon e^{i\theta}-i\chi)\sin^{\beta}(-i\epsilon e^{i\theta})} = \frac{\sin(\lambda\chi-\phi')}{\sinh^{\alpha}(\chi)\epsilon^{\beta-1}}e^{-i\frac{\pi}{2}} \int_{\pi}^{0}d\theta e^{-i(\beta-1)\theta} + \mathcal{O}(\epsilon^{2-\beta}) \nonumber\\
& = \frac{\sin(\lambda\chi-\phi')}{\sinh^{\alpha}(\chi) ~\epsilon^{\beta-1}} \frac{1}{(\beta-1)} \left[ 1- e^{-i(\beta-1)\pi} \right] + \mathcal{O}(\epsilon^{2-\beta}).
\end{align}
In the second-to-last equality, we make use of the series expansion for small $\epsilon$. Finally, on the last segment, we have
\begin{align}
\int^{\infty}_{\chi+\epsilon}d\tau \frac{\sin \left(\lambda \tau-\phi'\right)} {\sin^{\alpha}(-i\tau) \sin^{\beta}\left[-i(\tau-\chi)\right]} = \int_{\chi+\epsilon}^{\infty} d\tau \frac{(-i)^{-\alpha}(-i)^{-\beta}\sin(\lambda\tau - \phi')}{\sinh^{\alpha}(\tau)\sinh^{\beta}(\tau-\chi)} = - \int_{\chi+\epsilon}^{\infty} d\tau \frac{\sin(\lambda\tau - \phi')}{\sinh^{\alpha}(\tau)\sinh^{\beta}(\tau-\chi)}.
\end{align} 
Since the interference contribution to the tunneling current is given by the difference between the two integrals, we find
\begin{align}
\mathbb{I}_1 - \mathbb{I}_2 &= \int_{\epsilon}^{\chi-\epsilon} d\tau \frac{e^{i(1-\alpha)\pi}\sin(\lambda\tau-\phi')}{\sinh^{\alpha}(\tau)\sinh^{\beta}(\chi-\tau)} - \int_{\epsilon}^{\chi-\epsilon} d\tau \frac{e^{-i(1-\alpha)\pi}\sin(\lambda\tau-\phi')}{\sinh^{\alpha}(\tau)\sinh^{\beta}(\chi-\tau)} \nonumber\\
&~~+ \frac{\sin(\lambda\chi-\phi')}{\sinh^{\alpha}(\chi) ~\epsilon^{\beta-1}} \frac{1}{(\beta-1)} \left[ 1- e^{i(\beta-1)\pi} \right] - \frac{\sin(\lambda\chi-\phi')}{\sinh^{\alpha}(\chi) ~\epsilon^{\beta-1}} \frac{1}{(\beta-1)} \left[ 1- e^{-i(\beta-1)\pi} \right] + \mathcal{O}(\epsilon^{2-\beta}) \nonumber \\
&= 2i \sin((1-\alpha)\pi) \int_{\epsilon}^{\chi-\epsilon} d\tau \frac{\sin(\lambda\tau-\phi')}{\sinh^{\alpha}(\tau)\sinh^{\beta}(\chi-\tau)} - 2i \sin((\beta-1)\pi)\frac{\sin(\lambda\chi-\phi')}{\sinh^{\alpha}(\chi) } \frac{1}{\epsilon^{\beta-1}(\beta-1)}+ \mathcal{O}(\epsilon^{2-\beta}). 
\end{align}
We notice that the first integral has a singularity at $\chi$ when we take the $\epsilon \rightarrow 0$ limit. However, the singularity of the second term cancels the first. We can see this by writing
\begin{align}
\frac{\epsilon^{-(\beta-1)}}{(\beta-1)} = \frac{1}{(\beta-1) (\chi-\epsilon)^{\beta-1}} + \int_{\epsilon}^{\chi-\epsilon} d\tau \frac{1}{(\chi-\tau)^{\beta}}.
\end{align}
We also note that $\beta-1 = 1-\alpha$. Thus, we have
\begin{align}
\mathbb{I}_1 - \mathbb{I}_2 &= 2i \sin((\beta-1)\pi) \int_{\epsilon}^{\chi-\epsilon} d\tau \left[ \frac{\sin(\lambda\tau-\phi')}{\sinh^{\alpha}(\tau)\sinh^{\beta}(\chi-\tau)} - \frac{\sin(\lambda\chi-\phi')}{\sinh^{\alpha}(\chi)(\chi-\tau)^{\beta}} \right] - \frac{2i\sin((\beta-1)\pi)}{(\beta-1)\sinh^{\alpha}(\chi)}\frac{\sin(\lambda\chi-\phi')}{(\chi-\epsilon)^{\beta-1}} + \mathcal{O}(\epsilon^{2-\beta}).
\end{align}
We can now take the $\epsilon\rightarrow 0$ limit safely to obtain
\begin{align}
\mathbb{I}_1 - \mathbb{I}_2 &= -2i \sin((\beta-1)\pi) \int_{0}^{\chi} d\tau \left[ \frac{\sin(\phi'-\lambda\tau)}{\sinh^{\alpha}(\tau)\sinh^{\beta}(\chi-\tau)} - \frac{\sin(\phi'-\lambda\chi)}{\sinh^{\alpha}(\chi)(\chi-\tau)^{\beta}} \right] + \frac{2i\sin((\beta-1)\pi)}{(\beta-1)\sinh^{\alpha}(\chi)}\frac{\sin(\phi'-\lambda\chi)}{\chi^{\beta-1}}.
\end{align}
With the above result, we express the interference contribution to the tunneling current in terms of the function $F^{\text{int}}_{k}(\lambda_1,\lambda_2,\chi)$ as
\begin{align}
\langle I_T \rangle^{\text{int}} = - \frac{2e^{*2}V}{h} \frac{4\pi^2|\Gamma_1\Gamma_2|}{v_1^{\alpha}v_2^{\beta}\hbar^2} F_{k}^{\text{int}}\left[ \frac{e^* V_1}{\pi k_B \Theta }, \frac{e^* V_2}{\pi k_B \Theta},\left( \frac{l}{v_2}-\frac{l+a}{v_1} \right) \frac{\pi k_B \Theta}{\hbar} \right],
\end{align}
where we defined the function $F_{k}^{\text{int}}(\lambda_1,\lambda_2,\chi)$ as
\begin{align}
F_{k}^{\text{int}}(\lambda_1,\lambda_2,\chi) &= \frac{\sin\left( \frac{2\pi}{2k+1} \right)}{\pi(\lambda_1-\lambda_2)} \Bigg[ \int_{0}^{\chi} d\tau \left( \frac{\sin(\varphi - \lambda_2 \chi - (\lambda_1-\lambda_2)\tau)}{\sinh^{\frac{2k-1}{2k+1}}(\tau)\sinh^{\frac{2k+3}{2k+1}}(\chi-\tau)} - \frac{\sin(\varphi - \lambda_1 \chi)}{\sinh^{\frac{2k-1}{2k+1}}(\chi)(\chi-\tau)^{\frac{2k+3}{2k+1}}} \right) \nonumber \\ 
&~~~~~~~~~~~~~~~~~~~~~~~~~~~~~~~~~~~~~~~~~~~ -  \frac{2k+1}{2\sinh^{\frac{2k-1}{2k+1}}(\chi)}\frac{\sin(\varphi-\lambda_1\chi)}{\chi^{\frac{2}{2k+1}}} \Bigg]
\end{align}
and use the relation $\phi'-\lambda\tau = \varphi - \lambda_2 \chi - (\lambda_1-\lambda_2)\tau $. Note that the above function depends on the electric potentials $V_{1,2}$ applied to the two modes. However, in the $\chi\rightarrow 0$ limit, we can see that only  the potential difference matters. It can be checked both numerically and analytically that the limit of equal propagation times gives
\begin{align}
\lim_{\chi\rightarrow 0} F^{\text{int}}_{k}(\lambda_1,\lambda_2,\chi) = \cos\varphi. %
\end{align}
An analytical check requires some care. The integral contains expressions, which diverge in the limit of $\tau\rightarrow\chi$. One can put them under control by expressing $1/(\chi-\tau)^\beta$ and the most divergent part of $1/\sinh^\beta(\chi-\tau)$ as a derivative
$\frac{d}{(1-\beta)d\delta}(\chi+\delta-\tau)^{1-\beta}|_{\delta=0}$. Next, the resulting integral can be expressed in terms of the Euler $B$-function, which cancels the sine in front of the integral.
Thus, we have
\begin{align}
\lim_{lv_1\rightarrow(l+a)v_2}  F_{k}^{\text{int}}\left[ \frac{e^* V_1}{\pi k_B \Theta }, \frac{e^* V_2}{\pi k_B \Theta},\left( \frac{l}{v_2}-\frac{l+a}{v_1} \right) \frac{\pi k_B \Theta}{\hbar} \right] =  \cos\varphi,
\end{align}
as discussed in the main text.

 \section{Integral for thermal transport with a single constriction}\label{appendix:J_special}

We now evaluate the integral
\begin{align}
\mathbb{I} &\equiv \int_{-\infty}^{\infty} d\tau  \frac{\cos \left[  \epsilon+i\tau \right] }{\sin^{\beta+1}\left[ \epsilon+i\tau  \right]\sin^{\alpha}\left[ n(\epsilon+i \tau)  \right]}
\end{align}
for $\alpha+\beta=2$. We find that
\begin{align}
\mathbb{I}&= \int_{-\infty}^{-\epsilon} d\tau \frac{\cos ( i\tau) }{\sin^{\beta+1}( i\tau )\sin^{\alpha}( i n\tau )} +  \int_{\mathcal{C}} d\tau \frac{\cos ( i\tau ) }{\sin^{\beta+1}( i\tau )\sin^{\alpha}(i n\tau  )} +  \int_{\epsilon}^{\infty} d\tau \frac{\cos ( i\tau ) }{\sin^{\beta+1}( i\tau  )\sin^{\alpha}( i n\tau  )} \nonumber\\
&= -i\int_{-\infty}^{-\epsilon} d\tau \frac{\cosh ( \tau ) }{\sinh^{\beta+1}( -\tau  )\sinh^{\alpha}( -n\tau  )} +  \int_{-\pi}^{0} d\theta  \frac{ i\epsilon e^{i\theta} \cos (i \epsilon e^{i\theta}  ) }{\sin^{\beta+1}(i\epsilon e^{i\theta}  )\sin^{\alpha}( i n\epsilon e^{i\theta} )} +  i\int_{\epsilon}^{\infty} d\tau \frac{\cosh ( \tau ) }{\sinh^{\beta+1}(\tau)\sinh^{\alpha}( n\tau )} \nonumber\\
&=  \int_{-\pi}^{0} d\theta  \frac{ i\epsilon e^{i\theta} \cos ( i\epsilon e^{i\theta}  ) }{\sin^{\beta+1}(i\epsilon e^{i\theta}  )\sin^{\alpha}( i n\epsilon e^{i\theta} )} = -\frac{1}{n^{\alpha}\epsilon^{2}} \int_{-\pi}^{0} d\theta e^{-i2\theta} + \frac{\alpha}{6} \left( n^{\beta} - \frac{1}{n^{\alpha}} \right) \int_{-\pi}^{0} d\theta + \mathcal{O}(\epsilon)\nonumber\\
&\xrightarrow{\epsilon\rightarrow 0} \frac{\alpha\pi}{6} \left( n^{\beta} - \frac{1}{n^{\alpha}} \right),
\end{align}
where $\mathcal{C}$ is a semicircle contour in the lower half-plane around $\tau=0$ and we made use of the expansion
\begin{align}
\frac{\cos(z)}{\sin^{\beta+1}(z)\sin^{\alpha}(nz)} = \frac{1}{n^{\alpha}} \frac{1}{z^3} + \frac{\alpha}{6} \left[  n^{\beta} - \frac{1}{n^{\alpha}} \right] \frac{1}{z} + \mathcal{O}(z)~~~~ \text{for}~~~ \alpha+\beta=2.
\end{align}

\section{Integrals for thermal transport}\label{appendix:J_general}

In this Appendix, we solve two integrals with non-zero $\chi$. A general expression for the interference contribution to the tunneling heat current is 
\begin{align}
\langle J_Q \rangle^{\text{int}} = -\frac{\beta|\Gamma_1\Gamma_2|}{\hbar} \frac{2\cos\varphi}{v_1^{\alpha}v_2^{\beta}} \left( \frac{\pi k_B \Theta_2}{\hbar} \right)^{2} n^{\alpha} ~ \left( \mathcal{I}_1 + \mathcal{I}_2 \right),
\end{align}
where $\mathcal{I}_{1,2}$ are the integrals we need to solve; see equations (\ref{EqD2}) and (\ref{EqD10}). Since there are two branch points on the real axis for each integral, we use the same contours as in the electrical conductance case (Fig.~\ref{fig:contour-3} and Fig.~\ref{fig:contour-4}). We start with
\begin{align}
\label{EqD2}
\mathcal{I}_1&=\int_{-\infty}^{\infty} d\tau \frac{\cos[\epsilon + i (\tau-\chi)]}{\sin^{\alpha}[n(\epsilon + i \tau)]\sin^{\beta+1}[\epsilon + i (\tau-\chi)]} \nonumber\\
&= \left[ \int_{-\infty}^{-\epsilon}d\tau + \int_{\mathcal{C}_1}d\tau + \int_{\epsilon}^{\chi-\epsilon}d\tau + \int_{\mathcal{C}_2}d\tau + \int_{\chi+\epsilon}^{\infty}d\tau   \right] \frac{\cos[i (\tau-\chi)]}{\sin^{\alpha}[n i \tau]\sin^{\beta+1}[i (\tau-\chi)]}.
\end{align}
We simplify the integral on each of the segments
$(-\infty,-\epsilon)$, $\mathcal{C}_1$, $(+\epsilon,\chi-\epsilon)$, $\mathcal{C}_2$, and $(\chi+\epsilon,+\infty)$. First,
\begin{align}
\int_{-\infty}^{-\epsilon}d\tau \frac{\cos[i (\tau-\chi)]}{\sin^{\alpha}[ni \tau]\sin^{\beta}[i (\tau-\chi)]} = \int_{-\infty}^{-\epsilon} d\tau \frac{ (-i)^{-\beta-1}(-i)^{-\alpha}\cosh(\chi-\tau)}{\sinh^{\alpha}(-n\tau)\sinh^{\beta+1}(\chi-\tau)} = \int_{\epsilon}^{\infty} d\tau \frac{e^{-i\frac{\pi}{2}} ~\cosh(\chi+\tau)}{\sinh^{\alpha}(n\tau)\sinh^{\beta+1}(\chi+\tau)},
\end{align}
where we used $\alpha+\beta=2$. Next we consider the contour $\mathcal{C}_1$ defined as a semicircle passing through the lower half-plane around $\tau=0$,
\begin{align}
\int_{\mathcal{C}_1}d\tau \frac{\cos[i (\tau-\chi)]}{\sin^{\alpha}[ni \tau]\sin^{\beta+1}[i (\tau-\chi)]} = \int_{-\pi}^{0}d\theta \frac{i\epsilon e^{i\theta}\cos[i(\epsilon e^{i\theta}-\chi)]}{\sin^{\alpha}[in\epsilon e^{i\theta}]\sin^{\beta+1}[i(\epsilon e^{i\theta}-\chi)]} \sim \epsilon^{1-\alpha} + \mathcal{O}(\epsilon^{2-\alpha})\xrightarrow{\epsilon\rightarrow 0} 0,
\end{align}
where we used the following series expansion around $z=0$:
\begin{align}
\frac{\cos(z-w)}{\sin^{\alpha}(nz)\sin^{\beta+1}(z-w)}\Bigg|_{z=0}= \frac{\cos(w)}{n^{\alpha}\sin^{\beta+1}(-w)}\frac{1}{z^{\alpha}} + \mathcal{O}(z^{1-\alpha})~~~\text{for}~~~ \alpha+\beta=2.
\end{align}
Next, we find
\begin{align}
\int_{\epsilon}^{\chi-\epsilon}d\tau \frac{\cos[i (\tau-\chi)]}{\sin^{\alpha}[ni \tau]\sin^{\beta+1}[i (\tau-\chi)]} = \int_{\epsilon}^{\chi-\epsilon}d\tau \frac{e^{i\frac{\pi}{2}(\beta-\alpha+1)}\cosh(\chi-\tau)}{\sinh^{\alpha}(n\tau)\sinh^{\beta+1}(\chi-\tau)}.
\end{align}
Then we consider the contour $\mathcal{C}_2$ defined as a semicircle passing through the lower half-plane around $\tau=\chi$, and
\begin{align}
\int_{\mathcal{C}_2}d\tau \frac{\cos[i (\tau-\chi)]}{\sin^{\alpha}[ni \tau]\sin^{\beta+1}[i (\tau-\chi)]} &= \int_{-\pi}^{0}d\theta \frac{i\epsilon e^{i\theta} \cos[i\epsilon e^{i\theta}]}{\sin^{\alpha}[in\epsilon e^{i\theta}+in\chi]\sin^{\beta+1}[i\epsilon e^{i\theta}]} \nonumber\\
&= \int_{-\pi}^{0}d\theta \left[ \frac{(i\epsilon e^{i\theta})^{-\beta}}{\sin^{\alpha}(in\chi)} - \frac{\alpha n\cos(in\chi)(i\epsilon e^{i\theta})^{1-\beta}}{ \sin^{\alpha+1}(in\chi)} + \mathcal{O}(\epsilon^{2-\beta}) \right] \nonumber\\
& = \frac{i^{-(\alpha+\beta)}\epsilon^{-\beta}}{\sinh^{\alpha}(n\chi)} \int_{-\pi}^{0} d\theta e^{-i\beta\theta} - \frac{i^{-\beta}i^{-\alpha}n\alpha\cosh(n\chi)}{\sinh^{\alpha+1}(n\chi)} \epsilon^{1-\beta} \int_{-\pi}^{0}d\theta e^{i(1-\beta)\theta} + \mathcal{O}(\epsilon^{2-\beta}) \nonumber\\
&= \frac{i^{-(\alpha+\beta)}\epsilon^{-\beta}}{\sinh^{\alpha}(n\chi)} \frac{\left( 1-e^{i\beta\pi} \right)}{(-i)\beta} -  \frac{i^{-(\alpha+\beta)}n\alpha\cosh(n\chi)}{\sinh^{\alpha+1}(n\chi)} \epsilon^{1-\beta} \frac{\left( 1- e^{i(\beta-1)\pi} \right)}{(-i)(\beta-1)} + \mathcal{O}(\epsilon^{2-\beta}) \nonumber\\
= \frac{\epsilon^{-\beta}}{\beta\sinh^{\alpha}(n\chi)}&\left[ e^{-i\frac{\pi}{2}} - e^{i\frac{\pi}{2}(\beta-\alpha+1)} \right] - \frac{n\alpha\cosh(n\chi)\epsilon^{1-\beta}}{(\beta-1)\sinh^{\alpha+1}(n\chi)} \left[ e^{-i\frac{\pi}{2}} - e^{i\frac{\pi}{2}(\beta-\alpha-1)} \right] + \mathcal{O}(\epsilon^{2-\beta}),
\end{align}
where we use the following series expansion around $z=0$: 
\begin{align}
\frac{\cos(z)}{\sin^{\alpha}[n(z+w)]\sin^{\beta+1}(z)}\Bigg|_{z=0} = \frac{1}{\sin^{\alpha}(nw)z^{\beta+1}} - \frac{n\alpha\cos(n w)}{\sin^{\alpha+1}(n w)z^{\beta}} + \mathcal{O}(z^{1-\beta}).%
\end{align}
The first two  terms contribute to the integral since %
$1<\beta<2$.
Finally, on the last segment we find
\begin{align}
\int_{\chi+\epsilon}^{\infty}d\tau \frac{\cos[i (\tau-\chi)]}{\sin^{\alpha}[ni \tau]\sin^{\beta+1}[i (\tau-\chi)]} &= \int_{\chi+\epsilon}^{\infty} d\tau \frac{i^{-\beta-1}i^{-\alpha}\cosh(\tau-\chi)}{\sinh^{\alpha}(n\tau)\sinh^{\beta+1}(\tau-\chi)} = \int_{\chi+\epsilon}^{\infty} d\tau \frac{e^{i\frac{\pi}{2}}~\cosh(\tau-\chi)}{\sinh^{\alpha}(n\tau)\sinh^{\beta+1}(\tau-\chi)}.
\end{align}
We repeat the calculations with the second integral
\begin{align}
\label{EqD10}
\mathcal{I}_2&=\int_{-\infty}^{\infty} d\tau \frac{\cos[\epsilon - i (\tau-\chi)]}{\sin^{\alpha}[n(\epsilon - i \tau)]\sin^{\beta+1}[\epsilon - i (\tau-\chi)]} \nonumber\\
&= \left[ \int_{-\infty}^{-\epsilon}d\tau + \int_{\mathcal{C}_3}d\tau + \int_{\epsilon}^{\chi-\epsilon}d\tau + \int_{\mathcal{C}_4}d\tau + \int_{\chi+\epsilon}^{\infty}d\tau   \right] \frac{\cos[-i (\tau-\chi)]}{\sin^{\alpha}[-n i \tau]\sin^{\beta+1}[-i (\tau-\chi)]}.
\end{align}
Again, treating each of the segments separately, we find
\begin{align}
\int_{-\infty}^{-\epsilon}d\tau \frac{\cos[-i (\tau-\chi)]}{\sin^{\alpha}[-n i \tau]\sin^{\beta+1}[-i (\tau-\chi)]} = \int_{-\infty}^{-\epsilon}d\tau \frac{i^{-\alpha}i^{-\beta-1}\cosh(\chi-\tau)}{\sinh^{\alpha}[-n \tau]\sinh^{\beta+1} (\chi-\tau)} = \int^{\infty}_{\epsilon}d\tau \frac{e^{i\frac{\pi}{2}}~\cosh(\chi+\tau)}{\sinh^{\alpha}(n \tau)\sinh^{\beta+1} (\chi+\tau)}.
\end{align}
Next, we have the contour $\mathcal{C}_3$ defined as a semicircle passing through the upper half-plane around $\tau=0$, and
\begin{align}
\int_{\mathcal{C}_3}d\tau \frac{\cos[-i (\tau-\chi)]}{\sin^{\alpha}[-n i \tau]\sin^{\beta+1}[-i (\tau-\chi)]} = \int_{\pi}^{0}d\theta \frac{i\epsilon e^{i\theta} \cos[-i(\epsilon e^{i\theta}-\chi)] }{\sin^{\alpha}[-in\epsilon e^{i\theta}]\sin^{\beta+1}[-i(\epsilon e^{i\theta}-\chi)]} \sim \epsilon^{1-\alpha} + \mathcal{O}(\epsilon^{2-\alpha}) \xrightarrow{\epsilon\rightarrow 0} 0.
\end{align}
Next we get
\begin{align}
\int^{\chi-\epsilon}_{\epsilon}d\tau \frac{\cos[-i (\tau-\chi)]}{\sin^{\alpha}[-n i \tau]\sin^{\beta+1}[-i (\tau-\chi)]} = \int^{\chi-\epsilon}_{\epsilon}d\tau \frac{e^{-i\frac{\pi}{2}(\beta-\alpha+1)}\cosh(\chi-\tau)}{\sinh^{\alpha}(n\tau)\sinh^{\beta+1} (\chi-\tau)}.
\end{align}
Then we consider the contour $\mathcal{C}_4$ defined as a semicircle passing through the upper half-plane around $\tau=\chi$, and
\begin{align}
&\int_{\mathcal{C}_4}d\tau \frac{\cos[-i (\tau-\chi)]}{\sin^{\alpha}[-n i \tau]\sin^{\beta+1}[-i (\tau-\chi)]} = \int_{\pi}^{0} d\theta \frac{i\epsilon e^{i\theta}~\cos(-i\epsilon e^{i\theta})}{\sin^{\alpha}(-in\epsilon e^{i\theta}-in\chi)\sin^{\beta+1}(-i\epsilon e^{i\theta})}\nonumber \\
={}& \int_{\pi}^{0}d\theta ~i\epsilon e^{i\theta} \left[ \frac{(-i\epsilon e^{i\theta})^{-\beta-1}}{\sin^{\alpha}(-in\chi)} - \frac{\alpha n \cos(-in\chi)(-i\epsilon e^{i\theta})^{-\beta}}{\sin^{\alpha+1}(-in \chi)} + \mathcal{O}(\epsilon^{1-\beta}) \right]\nonumber \\
={}& \frac{(-i)^{-\beta-\alpha-1}i \epsilon^{-\beta}}{\sinh^{\alpha}(n\chi)}  \int_{\pi}^{0} d\theta e^{-i\beta\theta} - \frac{(-i)^{-\alpha-\beta-1}i\alpha n\cosh(n\chi)\epsilon^{1-\beta}}{\sinh^{\alpha+1}(n\chi)} \int_{\pi}^{0}d\theta e^{i(1-\beta)\theta} + \mathcal{O}(\epsilon^{2-\beta}) \nonumber\\
={}& \frac{(-i)^{-\beta-\alpha-1}i \epsilon^{-\beta}}{\sinh^{\alpha}(n\chi)} \frac{\left[1-e^{-i\beta\pi}\right]}{(-i)\beta} - \frac{(-i)^{-\alpha-\beta-1}\alpha n\cosh(n\chi)\epsilon^{1-\beta}}{\sinh^{\alpha+1}(n\chi)} \frac{\left[ 1-e^{i(1-\beta)\pi} \right]}
{(1-\beta)} + \mathcal{O}(\epsilon^{2-\beta}) \nonumber\\
={}& \frac{\epsilon^{-\beta}}{\beta\sinh^{\alpha}(n\chi)}  \left[ e^{i\frac{\pi}{2}} - e^{-i\frac{\pi}{2}(\beta-\alpha+1)} \right] - \frac{n\alpha \cosh(n\chi) \epsilon^{1-\beta}}{(\beta-1)\sinh^{\alpha+1}(n\chi)} \left[ e^{i\frac{\pi}{2}} - e^{-i\frac{\pi}{2}(\beta-\alpha-1)} \right] + \mathcal{O}(\epsilon^{2-\beta}).
\end{align}
Finally, on the last segment,
\begin{align}
 \int_{\chi+\epsilon}^{\infty}d\tau  \frac{\cos[-i (\tau-\chi)]}{\sin^{\alpha}[-n i \tau]\sin^{\beta+1}[-i (\tau-\chi)]} = \int_{\chi+\epsilon}^{\infty}d\tau  \frac{(-i)^{-\alpha}(-i)^{-\beta-1}\cosh(\tau-\chi)}{\sinh^{\alpha}(n \tau)\sinh^{\beta+1} (\tau-\chi)} =  \int_{\chi+\epsilon}^{\infty}d\tau  \frac{ e^{-i\frac{\pi}{2}} \cosh(\tau-\chi)}{\sinh^{\alpha}(n \tau)\sinh^{\beta+1} (\tau-\chi)}.
\end{align}
We now combine the two integrals and see that the contributions for several segments cancel. We find
\begin{align}
\mathcal{I}_1 + \mathcal{I}_2 &= \int_{-\infty}^{\infty} d\tau \frac{\cos[\epsilon + i (\tau-\chi)]}{\sin^{\alpha}[n(\epsilon + i \tau)]\sin^{\beta+1}[\epsilon + i (\tau-\chi)]}  + \int_{-\infty}^{\infty} d\tau \frac{\cos[\epsilon - i (\tau-\chi)]}{\sin^{\alpha}[n(\epsilon - i \tau)]\sin^{\beta+1}[\epsilon - i (\tau-\chi)]}\nonumber \\
&=  \int_{\epsilon}^{\chi-\epsilon}d\tau \frac{e^{i\frac{\pi}{2}(\beta-\alpha+1)}\cosh(\chi-\tau)}{\sinh^{\alpha}(n\tau)\sinh^{\beta+1}(\chi-\tau)} + \int^{\chi-\epsilon}_{\epsilon}d\tau \frac{e^{-i\frac{\pi}{2}(\beta-\alpha+1)}\cosh(\chi-\tau)}{\sinh^{\alpha}(n\tau)\sinh^{\beta+1} (\chi-\tau)} \nonumber \\
&~~+ \frac{\epsilon^{-\beta}}{\beta\sinh^{\alpha}(n\chi)} \left[ e^{-i\frac{\pi}{2}} - e^{i\frac{\pi}{2}(\beta-\alpha+1)} \right] - \frac{n\alpha\cosh(n\chi)\epsilon^{1-\beta}}{(\beta-1)\sinh^{\alpha+1}(n\chi)} \left[ e^{-i\frac{\pi}{2}} - e^{i\frac{\pi}{2}(\beta-\alpha-1)} \right] \nonumber \\
&~~+ \frac{\epsilon^{-\beta}}{\beta\sinh^{\alpha}(n\chi)} \left[ e^{i\frac{\pi}{2}} - e^{-i\frac{\pi}{2}(\beta-\alpha+1)} \right] - \frac{n\alpha \cosh(n\chi) \epsilon^{1-\beta}}{(\beta-1)\sinh^{\alpha+1}(n\chi)} \left[ e^{i\frac{\pi}{2}} - e^{-i\frac{\pi}{2}(\beta-\alpha-1)} \right] + \mathcal{O}(\epsilon^{2-\beta}).
\end{align}
These can be further simplified to 
\begin{align}
\mathcal{I}_1 + \mathcal{I}_2 &= 2\cos[\pi(\beta-\alpha+1)/2]\int_{\epsilon}^{\chi-\epsilon}d\tau \frac{\cosh(\chi-\tau)}{\sinh^{\alpha}(n\tau)\sinh^{\beta+1}(\chi-\tau)} - \frac{2\epsilon^{-\beta}\cos[\pi(\beta-\alpha+1)/2]}{\beta\sinh^{\alpha}(n\chi)} \nonumber \\
& ~~+ \frac{2n\alpha \cosh(n\chi)\epsilon^{1-\beta}\cos[\pi(\beta-\alpha-1)/2]}{(\beta-1)\sinh^{\alpha+1}(n\chi)} + \mathcal{O}(\epsilon^{2-\beta})\nonumber \\
&= 2\sin\left(\frac{\pi(\beta-\alpha)}{2}\right) \left[ \frac{\epsilon^{-\beta}}{\beta\sinh^{\alpha}(n\chi)} + \frac{n\alpha \cosh(n\chi)\epsilon^{1-\beta}}{(\beta-1)\sinh^{\alpha+1}(n\chi)} - \int_{\epsilon}^{\chi-\epsilon}d\tau \frac{\cosh(\chi-\tau)}{\sinh^{\alpha}(n\tau)\sinh^{\beta+1}(\chi-\tau)}  \right] + \mathcal{O}(\epsilon^{2-\beta}), 
\end{align}
where we used $\cos[\pi(\beta-\alpha\pm 1)/2]=\mp \sin[\pi(\beta-\alpha)/2]$. We can further simplify this by making use of the following formula:
\begin{align}
\frac{\epsilon^{-\mu}}{\mu} = \frac{(\chi-\epsilon)^{-\mu}}{\mu} + \int_{\epsilon}^{\chi-\epsilon}d\tau (\chi-\tau)^{-(\mu+1)} ~~~~~\text{for}~~ \mu\ne 0.
\end{align}
With this, the singularities in the integral cancel, and we may take the $\epsilon \rightarrow 0$ limit safely,
\begin{align}
\mathcal{I}_1 + \mathcal{I}_2& = 2\sin\left(\frac{\pi(\beta-\alpha)}{2}\right) \Bigg[ \frac{1}{\beta(\chi-\epsilon)^{\beta}\sinh^{\alpha}(n\chi)} + \frac{n\alpha \cosh(n\chi)}{(\beta-1)\sinh^{\alpha+1}(n\chi)(\chi-\epsilon)^{\beta-1}}\nonumber \\
&~~~+ \int_{\epsilon}^{\chi-\epsilon}d\tau \left( \frac{1}{(\chi-\tau)^{\beta+1}\sinh^{\alpha}(n\chi)} + \frac{n\alpha \cosh(n\chi)}{(\chi-\tau)^{\beta}\sinh^{\alpha+1}(n\chi)} - \frac{\cosh(\chi-\tau)}{\sinh^{\alpha}(n\tau)\sinh^{\beta+1}(\chi-\tau)} \right) \Bigg] + \mathcal{O}(\epsilon^{2-\epsilon}),\\
\lim_{\epsilon\rightarrow 0}(\mathcal{I}_1 + \mathcal{I}_2) &= 2\sin\left(\frac{\pi(\beta-\alpha)}{2}\right) \Bigg[ \frac{1}{\beta\chi^{\beta}\sinh^{\alpha}(n\chi)} + \frac{n\alpha \cosh(n\chi)}{(\beta-1)\sinh^{\alpha+1}(n\chi)\chi^{\beta-1}} \nonumber\\
& ~~~+ \int_{0}^{\chi}d\tau \left( \frac{1}{(\chi-\tau)^{\beta+1}\sinh^{\alpha}(n\chi)} + \frac{n\alpha \cosh(n\chi)}{(\chi-\tau)^{\beta}\sinh^{\alpha+1}(n\chi)} - \frac{\cosh(\chi-\tau)}{\sinh^{\alpha}(n\tau)\sinh^{\beta+1}(\chi-\tau)} \right) \Bigg]. 
\end{align}
Using the above result, we express the interference contribution to the tunneling heat current 
as
\begin{align}
\langle J_Q \rangle^{\text{int}} = \beta \frac{|\Gamma_1\Gamma_2|}{\hbar} \frac{(\pi k_B \Theta_2/\hbar)^2}{v_1^{\alpha}v_2^{\beta}} \frac{2\pi}{3} \frac{(2k-1)}{(2k+1)}\left(\frac{\Theta_2^2-\Theta_1^2}{\Theta_2^2}\right) ~F_{k}^{\text{int}}\left[ \frac{\Theta_1}{\Theta_2}, \frac{\pi k_B \Theta_2}{\hbar} \left( \frac{l}{v_2}-\frac{l+a}{v_1} \right) \right],
\end{align}
where the function $F_k^{\text{int}}(n,\chi)$ is defined as
\begin{align}
 F_k^{\text{int}}(n,\chi)&= -6 \frac{n^{\frac{2k-1}{2k+1}}}{ \alpha \pi (1-n^2) }\sin\left(\frac{2\pi}{2k+1}\right) \cos\varphi \Bigg[ \frac{1}{\beta\chi^{\beta}\sinh^{\alpha}(n\chi)} + \frac{n\alpha \cosh(n\chi)}{(\beta-1)\sinh^{\alpha+1}(n\chi)\chi^{\beta-1}} \nonumber \\
& ~~~+ \int_{0}^{\chi}d\tau \left( \frac{1}{(\chi-\tau)^{\beta+1}\sinh^{\alpha}(n\chi)} + \frac{n\alpha \cosh(n\chi)}{(\chi-\tau)^{\beta}\sinh^{\alpha+1}(n\chi)} - \frac{\cosh(\chi-\tau)}{\sinh^{\alpha}(n\tau)\sinh^{\beta+1}(\chi-\tau)} \right) \Bigg] .
\end{align}
We have checked analytically and numerically that the limit of equal propagation times works, {\it i.e.},
\begin{align}
\lim_{\chi\rightarrow 0} F^{\text{int}}_{k}(n,\chi) = \cos\varphi.
\end{align}
An analytical check is subtler than in Appendix \ref{appendix:I_general}. To put divergences under control, two tricks are needed. As in Appendix \ref{appendix:I_general}, divergent expressions should be rewritten as first or second derivatives of less divergent expressions with respect to a new infinitesimal variable. Another infinitesimal variable should be used to slightly shift the limits of integration so that all integrals in intermediate calculations converge. At the end, the infinitesimal variables are sent to zero.
As in Appendix \ref{appendix:I_general}, the sine factor in front of the integral is canceled by an Euler $B$-function. Thus, we obtain
\begin{align}
\lim_{lv_1 \rightarrow (l+a)v_2} F_k^{\text{int}} \left[ \frac{\Theta_1}{\Theta_2}, \frac{\pi k_B \Theta_2}{\hbar} \left( \frac{l}{v_2}-\frac{l+a}{v_1} \right) \right] = \cos\varphi,
\end{align}
as discussed in the main text. 

\section{Table of notations}\label{appendix:notations}

In this Appendix, we present the list of notations used in this paper. 

\noindent \begin{tabular}{l p{12.5cm}}
\hline 
\hline
$\nu_b$ & Filling fraction of the bulk of the incompressible liquid, $\nu_b=N/(2N+1)$ \\ 
$k = 1,2,\dots,N-1$ & Label for the incompressible states at $\nu=k/(2k+1)$\\
$\nu_1,\nu_2$ & Effective filling factors of the outer and inner modes\\ 
$\phi_1(x),\phi_2(x)$ & Bose fields describing the outer and inner chiral modes\\
$v_1,v_2$ & Velocities of the outer and inner modes\\
$\rho_1(x),\rho_2(x)$ & Charge densities for the outer and inner modes\\
$\varphi$ & Combination of the Aharonov-Bohm phase, statistical phase, and a non-universal phase difference between the constrictions\\
$e^*$ & Fundamental quasiparticle charge $e^*=e/(2k+1)$ \\
$e$ & Fundamental electric charge constant, $e>0$\\
$\epsilon$ & Ultraviolet cutoff\\
$\Theta_1,\Theta_2,\Theta$ & Temperatures of the outer and inner modes; the notation $\Theta$ is used if $\Theta_1=\Theta_2$\\
$\Psi_1,\Psi_2$ & Quasiparticle annihilation operators for the outer and inner modes\\
$\alpha$ & $\alpha=(2k-1)/(2k+1)$ reflects the scaling dimension of quasiparticles in the outer channel\\
$\beta$ & $\beta=(2k+3)/(2k+1)$ reflects the scaling dimension of quasiparticles in the inner edge channel\\
$T(x,y;t)$ & Tunneling operator, which transfers charge $e^*$ from the outer to inner mode\\
$T'(x,y;t)$ & Tunneling operator with a time derivative, $T'(x,y;t)\equiv \left(\partial_t \Psi^{\dagger}_2(y,t) \right) \Psi_1(x,t) $\\
$l$ & Distance between the two constrictions along the inner channel\\
$l+a$ & Distance between the two constrictions along the outer channel\\
$\Gamma_1,\Gamma_2$ & Tunneling amplitudes\\
$V_1,V_2$ & Electric potentials of  the outer and inner channels, $V\equiv V_1-V_2$\\
$\omega_q = e^*V/\hbar$ & Josephson frequency \\
$\theta(x,y)$ & Voltage-induced phase $\theta(x,y)\equiv e^* (V_1x/v_1-V_2y/v_2)/\hbar$\\
$I_T$ & Electric tunneling current operator\\
$J_Q$ & Thermal tunneling current operator\\
$G^{>}(x,y;t)$ & Tunneling correlations defined as $G^{>}(x,y;t)\equiv \langle T(x,y;t)T^{\dagger}(0,0;0) \rangle_0$\\
$G^{<}(x,y;t)$ & Tunneling correlations defined as $G^{<}(x,y;t)\equiv \langle T(0,0;0)T^{\dagger}(x,y;t) \rangle_0$\\
$\tilde{G}^{>}(\omega),\tilde{G}^{<}(\omega)$ & Fourier transform of Green's function $\tilde{G}^{\gtrless}(\omega)\equiv \int d\eta e^{-i\omega \eta} G^{\gtrless}(x,y;\eta)$\\
$\mathcal{G}_{1}^{>}(x,t),\mathcal{G}_{1}^{<}(x,t),\mathcal{G}_{2}^{>}(x,t),\mathcal{G}_{2}^{<}(x,t)$ & Thermal Green's functions defined for the outer and inner channels, Eqs. (\ref{eq:thermal_greens_functions},\ref{eq:t-g-f-2})\\
$\phi\equiv \varphi+\theta(l+a,l)$ & Combination of the voltage-induced phase $\theta(l+a,l)$ and $\varphi$\\
$\tau_1\equiv \pi k_B \Theta_2 (l+a)/\hbar v_1$ & Dimensionless propagation time between the two constrictions along the outer channel (expressed in terms of $\Theta_2$)\\
$\tau_2\equiv \pi k_B \Theta_2 l/\hbar v_2$ & Dimensionless propagation time between the two constrictions along the inner channel (expressed in terms of $\Theta_2$)\\
$\tau\equiv \pi k_B \Theta_2 t/\hbar$ & Dimensionless time at temperature $\Theta_2$\\
$\chi\equiv \tau_2-\tau_1$ & Dimensionless propagation time difference\\
$\lambda \equiv \hbar\omega_q/\pi k_B \Theta$ & Dimensionless Josephson frequency at temperature $\Theta$\\
$\phi'\equiv \phi - \lambda \tau_1$ & Shifted phase difference\\
$\lambda_1\equiv e^*V_1/\pi k_B \Theta,\lambda_2\equiv e^*V_2/\pi k_B \Theta$ & Ratio of the electric and thermal energy scales of the outer and inner channels at temperature $\Theta$\\
$n\equiv \Theta_1/\Theta_2$ & Ratio of the temperatures of the outer and inner channels\\
$\zeta_1,\zeta_2$ & Spatial coordinate rescaling factor $x\rightarrow \zeta_i x$ in the $i$th channel \\
$u$ & Propagation speed of the rescaled fields along both  channels\\
$\xi_1(x),\xi_2(x)$ & Bose fields describing the chiral modes defined in the basis of Eq. (\ref{eq:basis_change}) \\
$\psi_1(x),\psi_2(x)$ & Chiral fermion operators corresponding to the edge of a $\nu=2$ QH liquid \\
$t_1,r_1,t_2,r_2$ & Transmission and reflection coefficients at the two constrictions\\ 
$t= t_1t_2-r_1^*r_2$ & Effective transmission into drain D1, Fig. \ref{fig:3} \\
$r=t_1r_2+r_1t_2$ & Effective reflection into drain D2, Fig. \ref{fig:3}\\
\hline
\hline 
\end{tabular} 

\newpage
\twocolumngrid

\end{document}